\documentclass[11pt]{revtex4-1}

\usepackage{amsmath}
\usepackage{amssymb}
\usepackage{amsfonts}
\usepackage{graphicx}
\usepackage{bm}
\usepackage{hyperref}
\usepackage{color}
\usepackage{slashed}
\usepackage{braket}

\newcommand{\p}{{\bm{p}}}

\begin{document}

\title{Minimal-doubling and single-Weyl Hamiltonians}

\author{Tatsuhiro Misumi}
\email{misumi@phys.kindai.ac.jp}
\affiliation{Department of Physics, Kindai University, Higashi-Osaka, Osaka 577-8502, Japan}

\begin{abstract}
We develop a systematic Hamiltonian formulation of minimally doubled lattice fermions in $(3+1)$ dimensions, derive their nodal structures (structures of zeros), and classify their symmetry patterns for both four-component Dirac and two-component Weyl constructions. Motivated by recent single-Weyl proposals based on Bogoliubov-de Gennes (BdG) representation, we argue that the corresponding single-Weyl Hamiltonians are obtained from the minimal-doubling Hamiltonians supplemented by an appropriate species-splitting mass term, and we re-examine the non-onsite symmetry protecting the physical Weyl node in terms of a Ginsparg-Wilson-type relation. We then construct a one-parameter family of deformations that preserves all the symmetries and demonstrate that, once the parameter exceeds a critical value, additional Weyl nodes emerge and the system exits the single-node regime. This indicates that in interacting theories radiative corrections can generate symmetry-allowed counterterms, so maintaining the desired single-Weyl phase generically requires ``moderate" parameter tuning.
\end{abstract}

\maketitle
\tableofcontents
\newpage

\section{Introduction}

Formulating chiral fermions on a lattice is a long-standing problem at the intersection of quantum field theory, condensed-matter band theory, and nonperturbative gauge dynamics \cite{Wilson:1974sk, Creutz:1980zw}. 
The Nielsen-Ninomiya no-go theorem is most generally formulated in terms of conserved onsite charges with discrete spectra, together with locality, Hermiticity, and translation invariance. Under these standard assumptions, left- and right-handed gapless fermions must appear in matching numbers within each charge sector, thereby forbidding an isolated Weyl mode \cite{Karsten:1980wd, Nielsen:1980rz, Nielsen:1981xu, Nielsen:1981hk}.
Well-established lattice regularizations of vector-like gauge theories, such as Wilson fermions and staggered fermions, achieve practical control of doublers at the cost of either explicitly breaking chiral symmetry (breaking a conserved onsite charge) or introducing additional tastes \cite{Wilson:1975id,Kogut:1974ag, Susskind:1976jm, Kawamoto:1981hw,Sharatchandra:1981si,Golterman:1984cy,Golterman:1985dz,Kilcup:1986dg}. 
Domain-wall and overlap fermions realize single-flavor setups at the cost of breaking locality or onsite chiral symmetry \cite{Kaplan:1992bt, Shamir:1993zy, Furman:1994ky, Neuberger:1998wv, Ginsparg:1981bj, Luscher:1998pqa}.
The improved and combined approaches of Wilson and staggered fermions have also been proposed, including the generalized Wilson fermions \cite{Bietenholz:1999km,Creutz:2010bm,Durr:2010ch,Durr:2012dw, Misumi:2012eh,Cho:2013yha,Cho:2015ffa,Durr:2017wfi},
the staggered-Wilson fermions \cite{Adams:2009eb,Adams:2010gx,Hoelbling:2010jw, deForcrand:2011ak,Creutz:2011cd,Misumi:2011su,Follana:2011kh,deForcrand:2012bm,Misumi:2012sp,Misumi:2012eh,Durr:2013gp,Hoelbling:2016qfv,Zielinski:2017pko}, and the central-branch Wilson fermion \cite{Kimura:2011ik,Creutz:2011cd,Misumi:2012eh,Chowdhury:2013ux, Misumi:2019jrt, Misumi:2020eyx}.
By contrast, minimally doubled fermions realize the smallest allowed number of species within the no-go theorem while preserving an exact onsite chiral symmetry, typically by sacrificing full hypercubic invariance and other discrete symmetries \cite{Karsten:1981gd,Wilczek:1987kw,Creutz:2007af,Borici:2007kz,Bedaque:2008xs,Bedaque:2008jm, Capitani:2009yn,Kimura:2009qe,Kimura:2009di,Creutz:2010cz,Creutz:2010qm,Capitani:2010nn,Tiburzi:2010bm,Kamata:2011jn,Misumi:2012uu,Misumi:2012ky,Capitani:2013zta,Capitani:2013iha,Misumi:2013maa,Weber:2013tfa,Weber:2017eds,Durr:2020yqa}.

While minimally doubled fermions have primarily been discussed in Euclidean Lagrangian formulations, a Hamiltonian viewpoint is independently motivated. First, it is the natural language for real-time dynamics and for the classification of gapless topological band structures such as Weyl semimetals \cite{Wan:2010fyf}. Second, it is foundational for Hamiltonian lattice gauge theory \cite{Kogut:1974ag, Susskind:1976jm} and for emerging quantum-simulation approaches, where one often implements the dynamics directly in the operator formalism rather than through a path integral. Despite this situation, the Hamiltonian formulation of minimal-doubling fermions has not yet been investigated.

Another motivation for the present work arises from recent proposals aiming to isolate a single Weyl node by rewriting the Hamiltonian in a doubled Bogoliubov-de Gennes (BdG) or Nambu particle/hole representation and adding a momentum-dependent pairing term in the Hamiltonian formalism
\cite{Gioia:2025bhl}. Such models can possess exact, momentum-dependent conserved charges that commute with the lattice Hamiltonian while acting as a non-onsite generator in the position space. From the viewpoint of lattice chiral symmetry, these conserved non-onsite charges are naturally interpreted as Hamiltonian analogues of Ginsparg--Wilson-type chiral symmetries \cite{Creutz:2001wp, Fujikawa:2000my, Clancy:2023kla, Clancy:2023ino, Fidkowski:2023sif, Singh:2025sye}: they are conserved, but their spectra need not be quantized, and the associated ``chiral'' transformations are generically momentum- or position-dependent. Classification of the similar single-Weyl Hamiltonians are found in Ref.~\cite{Meyniel:2025euu}. For the attempt to define chiral fermions in staggered formulations and the related works, see \cite{Chatterjee:2024gje, Onogi:2025xir, Aoki:2025vtp}.
(See also Ref.~\cite{Eichten:1985ft} for the old related work.)

The first aim of this work is therefore methodological: we provide a unified Hamiltonian construction of minimal-doubling fermions and systematically identify which discrete symmetries are preserved or broken in the resulting Hamiltonians. 
The second aim is to clarify the relation between the minimal-doubling Hamiltonian and the recently proposed BdG-based single-Weyl Hamiltonians, and to analyze the latter in terms of their chiral structure and possible symmetry-allowed deformations. From the lattice-gauge-theory perspective, the key question is whether the single-node phase can be maintained in the interacting theory. Throughout this paper, we use the term “node” to mean a gapless point (equivalently, a zero of the Bloch Hamiltonian / energy spectrum) in the Brillouin zone.

In this paper, we formulate minimal-doubling lattice Hamiltonians in $(3+1)$ dimensions and investigate their discrete symmetry patterns while classifying the Hamiltonians based on the location of the zeros (nodes) and the discrete symmetries. Then, we re-examine BdG-based single-Weyl Hamiltonians within the minimal-doubling framework and analyze their non-onsite chiral symmetry in terms of the Ginsparg--Wilson-like relation. Building on these results, we introduce a symmetry-preserving one-parameter deformation of the single-Weyl Hamiltonian. A key result is that, although the deformation preserves all the symmetries, it can drive the system out of the single-node regime by generating additional Weyl nodes once the parameter exceeds a critical threshold. Since this deformation is symmetry-allowed, it can be generated in an interacting theory as a counterterm; consequently, maintaining the single-Weyl phase generically requires adjusting its coefficient.

This paper is organized as follows. In Sec.~\ref{sec:KW}, we provide a detailed derivation of the minimal-doubling Hamiltonians, derive their node structures and classify their symmetry patterns. In Sec.~\ref{sec:GT}, we discuss the relation of the single-Weyl and the minimal-doubling Hamiltonians, introduce a one-parameter deformation, and show that the deformation can produce additional Weyl nodes. Sec.~\ref{sec:CD} is devoted to the summary and discussion.

\section{Minimal-doubling lattice Hamiltonians}
\label{sec:KW}

The minimal-doubling (minimally doubled) fermion is one of the ultra-local lattice fermion setups, which yields two species of fermions and preserves one exact chiral symmetry.
As shown in Ref.~\cite{Creutz:2010cz}, the minimal-doubling Dirac operators in the Lagrangian formalism have four independent classes, including Karsten-Wilczek, Borici-Creutz, Twisted-ordering and Dropped-twisted-ordering fermions. In Appendix \ref{sec:KW-CKM-review}, we review the Lagrangian formalism of minimal-doubling fermions, with emphasis on the Karsten-Wilczek fermion.
Below, we will construct the $(3+1)$-dimensional Hamiltonian formalism of minimal-doubling fermions and investigate their discrete symmetries.

\subsection{Karsten-Wilczek type}
\label{sec:KWt}

The Karsten-Wilczek (KW) fermion \cite{Karsten:1981gd} is the prototype of minimal-doubling fermions, achieved by selecting a preferred spatial direction to lift the degeneracy of species. 

For exact chiral symmetry, the one-particle Hamiltonian $h$ (Bloch Hamiltonian) must commute with $\gamma_5$ as $[h,\gamma_5] = 0$. Any term proportional to $\beta = \gamma_0$ anticommutes with $\gamma_5$ and is forbidden. 
In the restricted class considered here, namely, Hamiltonians continuously connected to the standard massless Dirac Hamiltonian, we restrict ourselves to the $\alpha_j = \gamma_0 \gamma_j$ $(j=1,2,3)$ structures.


\subsubsection{Four-component fermion}

We first consider a cubic spatial lattice with sites ${\bm{x}}=(x,y,z)\in\mathbb{Z}^3$ and a four-component spinor
$\psi_{{\bm{x}}}$ at each site with the periodic boundary condition.
Gamma matrices $\gamma_{\mu}$ or $\alpha_{j} = \gamma_0 \gamma_{j}$ act on the Dirac spinor.
The position-space and momentum-space representations of the Dirac spinor have the relation
\begin{equation}
 \psi_{{\bm{x}}}
 \,=\, \int_{-\pi}^{\pi}\frac{d^3 p}{(2\pi)^3}\,
   e^{i{\bm{p}}\cdot{\bm{x}}}\,
   \psi_{{\bm{p}}}\,,
 \qquad
 {\bm{p}}\,=\,(p_1,p_2,p_3)\,,
\end{equation}
where we implicitly consider an infinite lattice volume and continuous momenta.

The massless and free Karsten-Wilczek (KW) Hamiltonian for a 4-component Dirac spinor is
\begin{align}
H_{\text{\tiny KW}}^{\text{\tiny D}}\,&=\, \int_{-\pi}^{\pi}\,\frac{d^3 p}{(2\pi)^3}\,
     \psi^\dagger_{{\bm{p}}}\,h^{\text{\tiny D}}_{\text{\tiny KW}}({\bm{p}})\,\psi_{{\bm{p}}}\,,
\label{eq:KW_Dirac}
\end{align}
with
\begin{align}
h^{\text{\tiny D}}_{\text{\tiny KW}}({\bm{p}}) \,&=\, \alpha_1 \sin p_1 + \alpha_2 \sin p_2 + \alpha_3 \,\mathcal{F}_{\text{\tiny KW}}({\bm{p}})\,,
\label{eq:KW_Dirac2}
\\
\mathcal{F}_{\text{\tiny KW}}({\bm{p}}) \,&=\, \sin p_3 + r \left( 2 - \cos p_1 - \cos p_2 \right).
\label{eq:KW_func}
\end{align}
Here, $r$ is a real parameter.
The position-space representation of the Hamiltonian is
\begin{align}
H_{\text{\tiny KW}}^{\text{\tiny D}}\,=\,\sum_{\bm x}\,\psi^\dagger_{\bm x}\,\Biggl[
  \sum_{j=1}^{3}\frac{\alpha_j}{2i}\bigl(\psi_{{\bm x}+{\bm e}_j}-\psi_{{\bm x}-{\bm e}_j}\bigr)
 +\alpha_3\Bigl(
 2\psi_{\bm x}-\sum_{k=1}^{2}\frac{\psi_{{\bm x}+{\bm e}_k}+\psi_{{\bm x}-{\bm e}_k}}{2}
 \Bigr)\Biggr]\,,
\end{align}
where ${\bm e}_j$ is the unit vector in the $j$ direction.
The energy dispersion of this fermion is $E^2({\bm{p}}) = \sin^2 p_1 + \sin^2 p_2 + \mathcal{F}_{\text{\tiny KW}}^2({\bm{p}})$. Vanishing energy requires $\sin p_1 = \sin p_2 = 0$, leading to four candidates in the momentum plane: $(p_1, p_2) \in \{(0,0), (\pi,0), (0,\pi), (\pi,\pi)\}$.
\begin{itemize}
    \item At $(0,0)$: $\mathcal{F}_{\text{\tiny KW}} =  \sin p_3$. This vanishes at $p_3=0$ and $p_3=\pi$. $\implies$ two gapless modes.
    \item At $(\pi,0)$ or $(0,\pi)$: $\mathcal{F}_{\text{\tiny KW}} = \sin p_3 + 2r$. For $|r| > 1/2$, these are never zeros, so gapped.
    \item At $(\pi,\pi)$: $\mathcal{F}_{\text{\tiny KW}}  = \sin p_3 + 4r$. For $|r| > 1/4$, these are never zeros, so gapped.
\end{itemize}
Thus, provided $|r| > 1/2$, the KW Hamiltonian of the Dirac spinor in (\ref{eq:KW_Dirac}) has two Dirac nodes strictly at ${\bm{p}}_A=(0,0,0)$ and ${\bm{p}}_B=(0,0,\pi)$, leading to two gapless Dirac modes.
Near ${\bm{p}}_A=(0,0,0)$, it behaves as $h \approx \bm{\alpha}\cdot{\bm{p}}$, while near ${\bm{p}}_B=(0,0,\pi)$, expanding $\sin(\pi+k_3) \approx -k_3$, we find $h \approx \alpha_1 k_1 + \alpha_2 k_2 - \alpha_3 k_3$.
The phase diagram on the number of species in the parameter space has been intensively studied in Ref.~\cite{Misumi:2012uu}, where the number of species becomes six when a coefficient of the dimension-3 counterterm exceeds a certain threshold. Although the study was performed in the Euclidean Lagrangian formalism, the similar phase structure is also expected to exist in the present Hamiltonian formalism.
This fact will be important in the discussion on the one-parameter deformation of single-Weyl Hamiltonian in Sec.~\ref{subsec:mu-deformation-and-tuning}.


Since the one-particle Hamiltonian depends only on $\alpha_j$, it strictly commutes with the chiral operator $\gamma_5$,
\begin{equation}
[h^{\text{\tiny D}}_{\text{\tiny KW}}, \,\gamma_5] = 0
\quad\implies \quad
e^{-i\gamma_5 \theta}\, h^{\text{\tiny D}}_{\text{\tiny KW}} 
\,e^{i\gamma_5 \theta}\,=\, h^{\text{\tiny D}}_{\text{\tiny KW}}\,,
\end{equation}
which means 
\begin{align}
&[H^{\text{\tiny D}}_{\text{\tiny KW}}, \,Q_5] \,=\, 0\,,
\\
&Q_5 \,=\, \sum_{{\bm x}}\psi^{\dagger}_{\bm x} \gamma_5 \psi_{\bm x} \,=\,  \int_{-\pi}^{\pi}\frac{d^3 p}{(2\pi)^3}\,
   \psi^{\dagger}_{\bm p} \,\gamma_5\,
   \psi_{{\bm{p}}}\,.
\end{align}
Thus, the many-body (field-theoretical) Hamiltonian possesses the $U(1)_A$ chiral symmetry, in addition to the $U(1)$ vector-like symmetry given by $[H^{\text{\tiny D}}_{\text{\tiny KW}},\, Q_0] = 0$ with $Q_0 = \sum_{{\bm x}}\psi^{\dagger}_{\bm x}\psi_{\bm x}$.

The term $\alpha_3(2-\cos p_1 - \cos p_2)$ in (\ref{eq:KW_func}) explicitly breaks the cubic rotational symmetry to its two-dimensional subgroup although the translational symmetry is automatically preserved.
Furthermore, some of the discrete symmetries are broken in the Hamiltonian:
Here, we adopt the standard implementations of P, T, C symmetries (one may choose any representation; for Dirac basis they are $U_P=\gamma^0$, $U_T=i\gamma^1\gamma^3$, $U_C=i\gamma^2 \gamma^0$) as
\begin{itemize}
\item
Parity symmetry:
\begin{equation}
  {\rm P}:\qquad U_P\,h(-\p)\,U_P^{-1}\,=\,h(\p)\,,
  \qquad U_P\,\alpha_i\, U_P^{-1}\,=\,-\alpha_i\,,
  \label{eq:P-def}
\end{equation}

\item
Time reversal symmetry:
\begin{equation}
  {\rm T}:\qquad U_T\,h(-\p)^{*}\,U_T^{-1}\,=\,h(\p)\,,
  \qquad U_T\,\alpha_i^{*}\, U_T^{-1}\,=\,-\alpha_i\,,
  \label{eq:T-def}
\end{equation}

\item
Charge conjugation invariance:
\begin{equation}
  {\rm C}:\qquad -U_C\,h(\p)^{T}\,U_C^{-1}\,=\,h(\p)\,,
  \qquad U_C \,\alpha_i^{T} \,U_C^{-1}\,=\,-\alpha_i\,.
  \label{eq:C-def}
\end{equation}
\end{itemize}
It is worth noting that this charge conjugation invariance is the relativistic field-theoretical definition, not the particle-hole symmetry (accompanied by momentum flipping $\p \to -\p$) relevant in the condensed-matter physics.
We here denote $s_i \equiv \sin p_{i}$, $c_i \equiv \cos p_{i}$ in the KW Hamiltonian \eqref{eq:KW_Dirac2}.
Then, for the Hamiltonian, one finds
\begin{itemize}

\item
P is broken: 
By applying \eqref{eq:P-def} to \eqref{eq:KW_Dirac2}, one finds
\begin{align}
  U_P \,h^{\text{\tiny D}}_{\text{\tiny KW}}(-\p)\, U_P^{-1}
  \,=\, \alpha_1 s_1+\alpha_2 s_2+\alpha_3[s_3-r(2- c_1 -c_2)]\,\neq\, h^{\text{\tiny D}}_{\text{\tiny KW}}(\p)\,,
\end{align}

\item
T is broken: 
Similarly, from \eqref{eq:T-def}, one finds
\begin{equation}
  U_T \,h^{\text{\tiny D}}_{\text{\tiny KW}}(-\p)^{*}\,U_T^{-1}
  \,=\,\alpha_1 s_1+\alpha_2 s_2+\alpha_3[s_3-r(2- c_1 -c_2)]\,\neq\, h^{\text{\tiny D}}_{\text{\tiny KW}}(\p)\,,
\end{equation}

\item
C is preserved: 
Using \eqref{eq:C-def}, one finds
\begin{align}
 - U_C \,h^{\text{\tiny D}}_{\text{\tiny KW}}(\p)^{T}\,U_C^{-1}
  \,=\, \alpha_1 s_1+\alpha_2 s_2+\alpha_3 [s_3+r(2- c_1 -c_2)] \,=\, h^{\text{\tiny D}}_{\text{\tiny KW}}(\p)\,.
\end{align}

\end{itemize}
Although P, T are each broken, the product PT is preserved: 
Both P and T flip $\p\to -\p$, so the combined action leaves $\p$ unchanged. 
Since PT is antiunitary, we may write its operation as $U_{PT}K$ with $K$ being the complex-conjugation operator.
Hence, at the level of single-particle matrices it acts as
$h(\p)\mapsto U_{PT}\,h(\p)^{*}\,U_{PT}^{-1}\,$.
Using that $h^{\text{\tiny D}}_{\text{\tiny KW}}(\p)$ has real coefficients and is a linear combination of $\alpha_i$, one finds
\begin{equation}
  {\rm PT}:\quad U_{PT}\,h^{\text{\tiny D}}_{\text{\tiny KW}}(\p)^{*}\,U_{PT}^{-1}\,=\,h^{\text{\tiny D}}_{\text{\tiny KW}}(\p)\,.
\end{equation}
Therefore, from the condensed-matter viewpoint, this model describes a phase analogous to a {\it PT-symmetric Dirac semimetal}. 
Since $h^{\text{\tiny D}}_{\text{\tiny KW}}$ is a $4\times 4$ Hamiltonian built from three mutually anti-commuting $\alpha$-matrices, its spectrum is $\pm E(p)$ with each energy level being two-fold degenerate. The PT symmetry is consistent with this structure, while the chiral symmetry gives the $E \leftrightarrow -E$ spectral symmetry.
In addition to the standard discrete symmetries, one can define the special parity-like symmetry, which can be regarded as a Hamiltonian version of mirror fermion symmetry discussed for the KW Dirac operator \cite{Weber:2017eds}
\begin{equation}
  \tilde {\rm P}:\qquad
  \p\mapsto (-p_1,-p_2,-p_3+\pi),\qquad
  \psi\mapsto \alpha_3\,\psi \,.
  \label{eq:modP-mom}
\end{equation}
Using $\alpha_3\alpha_{1,2}\alpha_3=-\alpha_{1,2}$,
one obtains the exact identity
\begin{equation}
  \alpha_3\,h^{\text{\tiny D}}_{\text{\tiny KW}}(-p_1,-p_2,-p_3+\pi)\,\alpha_3 \,=\, h^{\text{\tiny D}}_{\text{\tiny KW}}(p_1,p_2,p_3).
  \label{eq:modP-exact}
\end{equation}
Thus $\tilde {\rm P}$ is an exact symmetry of the single-particle Hamiltonian $h^{\text{\tiny D}}_{\text{\tiny KW}}(\p)$.


In lattice gauge theory with Lagrangian formalism of minimally doubled fermions, it has been argued that the breaking of discrete symmetries allows radiative corrections, leading to the necessity of tuning the parameters in the dimension-3 and -4 counterterms \cite{Bedaque:2008xs, Bedaque:2008jm, Capitani:2009yn, Capitani:2010nn, Capitani:2013zta, Weber:2013tfa, Weber:2017eds}.
In our Hamiltonian formalism, however, we have the combined PT symmetry, and it is an interesting question whether or not it prohibits the generation of dimension-3 counterterms.

\subsubsection{Two-component fermion}
\label{sec:KWw}

We now consider a two-component spinor
$\chi_{{\bm{x}}}$ at each site.
Pauli matrices $\sigma_{1,2,3}$ act on the Weyl spinor.
The position-space and momentum-space representations are related as $\chi_{\bm{x}}
 = \int_{-\pi}^{\pi}\frac{d^3 p}{(2\pi)^3}\,e^{i{\bm{p}}\cdot{\bm{x}}} \chi_{\bm p}$.
The massless and free KW Hamiltonian of the 2-component Weyl spinor is
\begin{align}
H_{\text{\tiny KW}}^{\text{\tiny W}}\,&=\, \int_{-\pi}^{\pi}\frac{d^3 p}{(2\pi)^3}\,
     \chi^\dagger_{\bm{p}}\,h^{\text{\tiny W}}_{\text{\tiny KW}}({\bm{p}})\,\chi_{\bm{p}}\,,
\label{eq:KW_Weyl}
\end{align}
with
\begin{align}
h^{\text{\tiny W}}_{\text{\tiny KW}}({\bm{p}}) = \sigma_1 \sin p_1 + \sigma_2 \sin p_2 + \sigma_3 \, \mathcal{F}_{\text{\tiny KW}}({\bm{p}})\,.
\label{eq:KW_W}
\end{align}
with $\sigma_j$ being the Pauli matrices and $\mathcal{F}_{\text{\tiny KW}}({\bm{p}})$ being defined \eqref{eq:KW_func}.
The position-space representation of the Hamiltonian is
\begin{align}
H_{\text{\tiny KW}}^{\text{\tiny W}}\,=\,\sum_{\bm x}\,\chi^\dagger_{\bm x}\,\Biggl[
  \sum_{j=1}^{3}\,\frac{\sigma_j}{2i}\,\bigl(\chi_{{\bm x}+{\bm e}_j}-\chi_{{\bm x}-{\bm e}_j}\bigr)
 \,+\,\sigma_3\,\Bigl(
 2\chi_{\bm x}-\sum_{k=1}^{2}\frac{\chi_{{\bm x}+{\bm e}_k}+\chi_{{\bm x}-{\bm e}_k}}{2}
 \Bigr)\Biggr]\,.
\end{align}
Again, provided $|r| > 1/2$, the KW Hamiltonian for Weyl lattice fermions in (\ref{eq:KW_Weyl}) contains two Weyl nodes strictly at ${\bm{p}}_A=(0,0,0)$ and ${\bm{p}}_B=(0,0,\pi)$, leading to the two gapless Weyl modes.
Near ${\bm{p}}_A={\bm 0}$, it behaves as $h \approx \bm{\sigma}\cdot{\bm{p}}$, while near ${\bm{p}}_B=(0,0,\pi)$ it behaves as $h \approx \sigma_1 p_1 + \sigma_2 p_2 - \sigma_3 p_3$.

The exact flavor-chiral symmetry possessed by this Weyl setup is the $U(1)$ onsite symmetry
\begin{equation}
[h^{\text{\tiny W}}_{\text{\tiny KW}},\, {\bm 1}_\sigma] \,=\, 0
\quad\implies \quad
e^{-i\theta} \,h^{\text{\tiny W}}_{\text{\tiny KW}}\, 
e^{i\theta}\,=\, h^{\text{\tiny W}}_{\text{\tiny KW}}\,,
\label{eq:KW_U1V}
\end{equation}
which means
\begin{align}
&[H^{\text{\tiny W}}_{\text{\tiny KW}}, \,Q_0] \,=\, 0\,,
\\
&Q_0 \,=\, \sum_{{\bm x}}\chi^{\dagger}_{\bm x} \chi_{\bm x} \,=\,  \int_{-\pi}^{\pi}\frac{d^3 p}{(2\pi)^3}\,
   \chi^{\dagger}_{\bm p}
   \chi_{\bm{p}}\,.
\end{align}
Here ${\bm 1}_\sigma$ denotes the $2 \times 2$ identity matrix acting on the Weyl-spinor space.
This Hamiltonian breaks the cubic rotational symmetry to the two-dimensional subgroup, and also breaks both Parity and Time-reversal invariances. 
In this sense, this Hamiltonian corresponds to the {\it P-broken magnetic Weyl semimetal} in the condensed-matter context.
In comparison to the case of the Dirac formalism, the Weyl KW formalism does not possess the combined PT invariance, while it still has the special parity-like symmetry $ \tilde {\rm P}:\,\p\mapsto (-p_1,-p_2,-p_3+\pi),\,\chi\mapsto \sigma_3\,\chi$.

Let us analyze the low-energy effective theory in this setup.
We have the two Weyl nodes in momentum space at ${\bm p}_A$ and ${\bm p}_B$. As with the case for the Lagrangian formalism introduced in Appendix \ref{sec:KW-CKM-review}, by using the approximate projection operator $P^A_\p = (1+\cos p_3)/2,\, P^B_\p = (1-\cos p_3)/2$ one can define point-split fields \cite{Creutz:2010qm, Creutz:2010bm} as
\begin{itemize}
    \item Sector A (${\bm{p}} \approx {\bm{p}}_A$): $\chi_R({\bm k}) \equiv P^{A}_{{\bm k} + \p_A}\chi_{{\bm k}+\p_A}$ obeys $i\partial_t \chi_R ({\bm k}) = \bm{\sigma}\cdot{\bm{k}} \,\chi_R({\bm k})$\,,
    \item Sector B (${\bm{p}} \approx {\bm{p}}_B$): $\chi_L({\bm k}) = \sigma_3 P^{B}_{{\bm k} + \p_B}\chi_{{\bm k}+\p_B}$ obeys $i\partial_t \chi_L ({\bm k}) = -\bm{\sigma}\cdot{\bm{k}}\, \chi_L ({\bm k})$\,.
\end{itemize}
These are regarded as Weyl equations with opposite chiralities (helicities). Thus, we can define a 4-component Dirac spinor
\begin{equation}
\Psi_{{\bm{x}}} \,\equiv\, \int\frac{d^3 k}{(2\pi)^3}e^{i {\bm k}\cdot {\bm x}} 
\begin{pmatrix} \chi_R ({\bm{k}}) \\ \chi_L ({\bm{k}}) \end{pmatrix}
\,=\, 
\frac{1}{2}\begin{pmatrix} \chi_{\bm{x}}+\frac{\chi_{\bm{x}+{\bm e}_3}+\chi_{\bm{x}-{\bm e}_3}}{2} \\ (-1)^{x_3} \sigma_3 \left(\chi_{\bm{x}}-\frac{\chi_{\bm{x}+{\bm e}_3}+\chi_{\bm{x}-{\bm e}_3}}{2}\right) \end{pmatrix}
\,,
\label{eq:DfW2}
\end{equation}
which means that the low-energy theory obtained by combining the two patches can be organized into a single massless Dirac fermion (a pair of Weyl fermions of opposite chirality).
It is, however, notable that the onsite $U(1)$ symmetry possessed by the Hamiltonian (\ref{eq:KW_U1V}) corresponds to the $U(1)$ vector symmetry of the single Dirac fermion composed of the two Weyl species \eqref{eq:DfW2}, while $U(1)_A$ chiral symmetry should be expressed as a non-onsite transformation, thus be broken even for the free theory in this setup. In the interacting theory with gauge field, this breaking of $U(1)_A$ is expected to yield the chiral anomaly. This point will later be important in the construction of the single-Weyl formulation in Sec.~\ref{sec:GT}.

As shown in Appendix \ref{sec:KW-CKM-review}, one can introduce the species-splitting mass term to the minimal-doubling fermion based on the flavor representation of point-split fields \cite{Creutz:2010bm}. Such a term gaps one of the two Weyl nodes while preserving the ordinary onsite fermion-number $U(1)$ symmetry. For instance, by introducing the species-splitting mass term into the KW two-component Hamiltonian \eqref{eq:KW_W}, one reaches a single-Weyl Hamiltonian
\begin{align}
h^{\text{\tiny split}}_{\text{\tiny KW}}({\bm{p}}) \,=\, \sigma_1 \sin p_1 + \sigma_2 \sin p_2 + \sigma_3 \mathcal{F}_{\text{\tiny KW}}({\bm{p}})
+ 1-\cos p_3 \,,
\label{eq:spKW}
\end{align}
which keeps the Weyl node at ${\bm p}_{A} = (0,0,0)$ massless while gaps the other at ${\bm p}_{B} = (0,0,\pi)$. 
In this setup, however, the onsite fermion-number $U(1)$ symmetry does not work to protect the remaining gapless node. Thus, although Eq.~\eqref{eq:spKW} yields a single gapless Weyl node in the free theory, it can be gapped due to the radiative correction in the interacting theory, thus it does not provide a symmetry-protected chiral single-Weyl construction.
In the chiral-symmetric Weyl setup discussed in Sec.~\ref{sec:GT}, the analogous species-splitting term is introduced not in flavor space but in the Bogoliubov-de Gennes (BdG) representation on the Nambu (particle/hole) space, where the onsite $U(1)$ is explicitly broken.



\subsection{Twisted-ordering type}
\label{sec:BC_TO}

In the following two subsections, we construct other classes of minimal-doubling Hamiltonians.
The first one is the twisted-ordering (TO) fermion \cite{Creutz:2010cz}, which is defined on an orthogonal lattice but utilizes a cyclic twist to maintain minimal doubling.
The twisted-ordering Hamiltonian for the Dirac spinor is 
\begin{align}
h^{\text{\tiny D}}_{\text{\tiny TO}}({\bm{p}}) &\,=\, \sum_{j=1}^3 \,\alpha_j \left[ \sin p_j +  \cos p_{j+1} -1 \right]
\nonumber\\
&\,=\,
\alpha_1 (\sin p_{1} + \cos p_{2} - 1 )
+
\alpha_2 (\sin p_{2} + \cos p_{3} - 1 )
+
\alpha_3 (\sin p_{3} + \cos p_{1} - 1 )\,,
\label{eq:TOh}
\end{align}
where we use the index $j+1$ as a mod-3 number, for instance $3+1 \to 1$.
This Hamiltonian has two nodes (zeros) at ${\bm{p}}=(0,0,0)$ and ${\bm{p}}=(\pi/2,\pi/2,\pi/2)$, yielding two gapless Dirac modes.
It possesses the $U(1)_A$ chiral symmetry in addition to the $U(1)$ vector symmetry as in the KW Dirac Hamiltonian in Sec.~\ref{sec:KWt}.
Regarding the discrete rotational symmetry, this model breaks cubic symmetry but, in this case, preserves the $Z_3$ subgroup.
Regarding P, T, C symmetries, the Hamiltonian breaks each of P and T, but preserves the combined PT symmetry while it does not have the special parity-like symmetry $\tilde{\rm P}$ possessed by the KW Dirac Hamiltonian.

The twisted-ordering Hamiltonian for Weyl spinors is
\begin{align}
h^{\text{\tiny W}}_{\text{\tiny TO}}({\bm{p}}) &\,=\, \sum_{j=1}^3 \,\sigma_j \left[ \sin p_j +  \cos p_{j+1} -1 \right]
\nonumber\\
&\,=\,
\sigma_1 (\sin p_{1} + \cos p_{2} - 1 )
+
\sigma_2 (\sin p_{2} + \cos p_{3} - 1 )
+
\sigma_3 (\sin p_{3} + \cos p_{1} - 1 )\,.
\label{eq:TOweyl}
\end{align}
It has two Weyl nodes at ${\bm{p}}_A=(0,0,0)$ and ${\bm{p}}_B=(\pi/2,\pi/2,\pi/2)$, yielding two gapless (massless) Weyl modes. 
They carry opposite physical chiralities (helicities), and therefore can be combined into a single massless Dirac fermion in the low-energy theory.
While it breaks P and T symmetries, it possesses the $U(1)$ onsite symmetry, which corresponds to $U(1)$ vector symmetry of the single Dirac fermion composed of the two Weyl modes at $\p_A, \p_B$ as in the KW Weyl Hamiltonian in Sec.~\ref{sec:KWw}.


\subsection{Borici-Creutz type}
\label{sec:BC}

The Borici-Creutz fermion \cite{Creutz:2007af, Borici:2007kz} places the two zeros along the main diagonal of the Brillouin zone, preserving an $S_3$ permutation subgroup of the cubic symmetry.
The Borici-Creutz Hamiltonian for Dirac spinors is 
\begin{equation}
h^{\text{\tiny D}}_{\text{\tiny BC}}({\bm{p}})\,=\, \sum_{j=1}^3\,[ \alpha_{j}\sin (p_{j} + \pi/4) \,
-\, \alpha_{j}'\sin (p_{j} - \pi/4) \,-\,\alpha_j]\,,
\label{eq:BCh}
\end{equation} 
with
\begin{align} 
\alpha'_{j} \,=\, A_{j i} \alpha_{i},
\quad\quad
A \,=\,  \left( 
\begin{matrix}
-1 & 1 & 1  \\
1 & -1 & 1  \\
1 & 1 & -1 
\end{matrix}
\right)\,.
\label{eq:matA}
\end{align}
The two nodes are located at ${\bm{p}}=(\pi/4,\pi/4,\pi/4)$ and ${\bm{p}}=-(\pi/4, \pi/4, \pi/4)$, giving two massless Dirac modes.
It preserves the $U(1)_A$ chiral symmetry and the $U(1)$ vector-like symmetry.
Regarding the discrete rotational symmetry, the model breaks the cubic symmetry but preserves the $S_3$ subgroup, which is the largest rotational symmetry in the known the minimal-doubling Hamiltonian. 
It breaks each of P and T, but preserves the combined PT symmetry as with the other minimal-doubling Hamiltonians, while it does not have the special parity-like symmetry.

The Borici-Creutz Hamiltonian for Weyl spinors is
\begin{align}
h^{\text{\tiny W}}_{\text{\tiny BC}}({\bm{p}})\,=\, \sum_{j}\,[ \sigma_{j}\sin (p_{j} + \pi/4) \,
-\, \sigma_{j}'\sin (p_{j} - \pi/4) -\sigma_j]\,,
\end{align}
with $\sigma'_{j} = A_{j i} \sigma_{i}$.
It has two Weyl nodes at ${\bm{p}}=(\pi/4,\pi/4,\pi/4)$ and ${\bm{p}}=-(\pi/4, \pi/4, \pi/4)$, giving two massless Weyl modes with the $U(1)$ onsite symmetry preserved and P, T symmetries broken.
Again, the two Weyl modes carry opposite physical chiralities, and therefore can be combined into a single massless Dirac fermion in the low-energy theory.

\subsection{Classification and Generalization}
\label{sec:CG}

In this subsection, we pursue classification and generalization of the minimal-doubling Hamiltonians.  
As in the Lagrangian formalism of minimal-doubling fermions in Ref.~\cite{Creutz:2010cz},
one can classify the expressions of minimal-doubling Hamiltonians into two types:

The first type is expressed as
\begin{equation} 
h_1\,=\, \sum_{j=1}^{3}\alpha_{j}\sin p_{j}\,
+\, r\sum_{i=2,3}\sum_{k=1,2} \alpha_{i}R_{ik}(\cos p_{k}-1)\,,
\label{GENE1}
\end{equation}
where we take $i=2,3$ and $k=1,2$ while $j=1,2,3$. 
This type breaks the cubic rotational symmetry severely.
For example, let us consider the following $R$
\begin{align}
R&=\left( 
\begin{matrix}
0\,\, & 0 \\
1\,\, & 1
\end{matrix}
\right)\,,
\label{RKW}
\\
R&=\left( 
\begin{matrix}
1\,\, & 0 \\
0\,\, & 1 
\end{matrix}
\right)\,.
\label{RT}
\end{align}
With (\ref{RKW}), the Hamiltonian (\ref{GENE1}) reduces to Karsten-Wilczek one in \eqref{eq:KW_Dirac2}.
With (\ref{RT}) ($r=1$), it reduces to the type, called the dropped
twisted-ordering fermion \cite{Creutz:2010cz},
\begin{align}
h_1 \,=\,   \alpha_{1} \sin p_{1}
+  \alpha_{2} ( \sin p_{2} + \cos p_{1} - 1 )
+ \alpha_{3} ( \sin p_{3} + \cos p_{2} - 1) \,,
\label{eq:DTO}
\end{align}
which has two gapless nodes at ${\bm p}=(0,0,0)(0,0,\pi)$.
Of course, we need to set $r$ so as to keep minimal-doubling. 

The second type is expressed as
\begin{equation}
h_2\,\,=\,\, \sum_{j=1}^{3}\,[ \alpha_{j}\sin (p_{j}+\beta_{j}) \,
-\, \alpha_{j}'\sin (p_{j}-\beta_{j}) ]\,-\, \alpha ,
\label{GENE2}
\end{equation} 
where $\alpha_{j}'=A_{ji}\alpha_{i}$.
Here $\beta_{j}$ and $\alpha$ are related with $A$ as
$\alpha=\sum_{j}\alpha_{j}\sin2\beta_{j}=\sum_{j}\alpha_{j}'\sin 2\beta_{j}$, which means that $\sin 2\beta_{j}$ is an eigenvector of $A$.  
Thus, once $A$ is fixed, $\beta_{j}$ and $\alpha$ are determined up to an overall factor.  
By imposing these conditions on $A$,
$\beta_{j}$ and $\alpha$, the Hamiltonian can be minimal-doubling, where $\pm\beta_{j}$ correspond to locations of two nodes.  
The Hamiltonian (\ref{GENE2}) reduces to Borici-Creutz one \eqref{eq:BCh} by
choosing \eqref{eq:matA} for $A$, 
which fixes $\beta_{j}=\pi/4$ and $\alpha=\sum_{j=1}^{3} \alpha_j$.
The Hamiltonian (\ref{GENE2}) also yields the twisted-ordering one by choosing $A$ as
\begin{align} 
A \,=\,  \left( 
\begin{matrix}
0 & 1 & 0 \\
0 & 0 & 1 \\
1 & 0 & 0
\end{matrix}
\right)\,, 
\label{GENE2-T}
\end{align}
which fixes $\beta_{j}=\pi/4$ and $\alpha=\sum_{j=1}^{3}\alpha_{j}$.
This agrees with the expression \eqref{eq:TOh} by shifting $p_{j} \to p_{j} -\pi/4$.
This type can maintain a relatively large subgroup of the cubic rotational symmetry such as $S_3$ or $Z_3$.

\section{Chiral-symmetric single-Weyl Hamiltonians}
\label{sec:GT}

We now turn to a single Weyl fermion formulation, which uses the BdG formalism to attempt to remove the doubler, leaving a single Weyl node, in the KW Hamiltonian \eqref{eq:KW_Dirac2}.
In this section, we first show the relation of the Karsten-Wilczek minimal-doubling Hamiltonian and the single-Weyl model proposed by Gioia and Thorngren \cite{Gioia:2025bhl}, and study its properties in terms of chiral symmetry.
Finally, we propose one-parameter deformation of the single-Weyl Hamiltonian, who keeps all the symmetries of the original Hamiltonian, and show that it can destabilize the single-Weyl situation, leading to potential necessity of tuning parameters in an interacting theory.

\subsection{Review of Gioia--Thorngren setup}
\label{sec:GT-review}

In this subsection we recall in some detail the construction of the single-Weyl model \cite{Gioia:2025bhl}, with focus on its relation to Karsten-Wilczek minimally doubled Hamiltonian in the previous section.

The starting point is a time-reversal-broken two-band tight-binding model on a cubic lattice, a magnetic Weyl semimetal, described by the second-quantized Hamiltonian.
It is completely equivalent to the Karsten-Wilczek minimal-doubling Weyl Hamiltonian in (\ref{eq:KW_W})
\begin{align}
H_{\text{\tiny KW}}^{\text{\tiny W}}\,&=\, \int_{-\pi}^{\pi}\frac{d^3 p}{(2\pi)^3}\,\chi^\dagger_{\bm{p}}\,h^{\text{\tiny W}}_{\text{\tiny KW}}({\bm{p}})\,\chi_{\bm{p}}\,,
\\
 h^{\text{\tiny W}}_{\text{\tiny KW}}({\bm{p}}) \,&=\, \sigma_1 \sin p_1 + \sigma_3 \sin p_2 + \sigma_2 \mathcal{F}_{\text{\tiny KW}}({\bm{p}})\,,
  \label{eq:H2-def}
\end{align}
with $\mathcal{F}_{\text{\tiny KW}}({\bm{p}}) = \sin p_3 + r \left( 2 - \cos p_1 - \cos p_2 \right)$.
Here, we have performed the appropriate unitary transformation on the original KW Hamiltonian in (\ref{eq:KW_W}) to match the notation to that in \cite{Gioia:2025bhl}.
As we have discussed in Sec.~\ref{sec:KW}, for $|r|>1/2$ the spectrum contains two Weyl nodes at
\begin{equation}
  {\bm{p}}_A=(0,0,0),\qquad
  {\bm{p}}_B=(0,0,\pi)\,,
\end{equation}
and the model has an onsite $U(1)$ symmetry generated by the charge
$Q_0 =\sum_{{\bm x}}\chi^{\dagger}_{\bm x} \chi_{\bm x}$,
which acts as $\chi_{\bm x}\mapsto e^{i\theta}\chi_{\bm x}$.

To obtain a single Weyl fermion in the IR, the above $U(1)$ symmetry must be broken in order to gap one of the nodes.
To implement the symmetry breaking in a convenient way, Ref.~\cite{Gioia:2025bhl} passes to a Bogoliubov--de Gennes (BdG) representation, where one introduces the Nambu spinor,
\begin{equation}
  \psi^{\text{\tiny BdG}}_{{\bm{p}}}
  \,=\, 
  \begin{pmatrix}
\chi_{\bm p} \\
\chi^{*}_{-{\bm p}}
  \end{pmatrix}\,,
  \qquad
  h^{\text{\tiny BdG}}_{\text{\tiny KW}}({\bm{p}}) \,=\, 
\frac{1}{2}
\begin{pmatrix} 
h^{\text{\tiny W}}_{\text{\tiny KW}}({\bm{p}}) & {\bm 0} \\ 
{\bm 0} & -h^{\text{\tiny W}}_{\text{\tiny KW}}(-{\bm{p}})^{*} \end{pmatrix}\,.
\label{eq:bdg}
\end{equation}
Note that this slightly differs from the conventional definition of the Nambu spinor such as 
$\psi^{\text{\tiny BdG}}_{{\bm{p}}} = ( \chi_{\bm p}, -i\sigma_2 \chi^{*}_{-{\bm p}} )^T$, 
but they are unitary-equivalent.
Then, one writes the single-particle BdG Hamiltonian \eqref{eq:bdg} as
\begin{equation}
  h^{\text{\tiny BdG}}_{\text{\tiny KW}}({\bm{p}})
  \,=\, \frac{1}{2}\Bigl[
     \sigma_1 \, \sin p_1
      +\sigma_3 \, \sin p_2
      + \tau_3\sigma_2 \, \sin p_3
      + \sigma_2 \, r \left( 2 - \cos p_1 - \cos p_2 \right)
    \Bigr],
  \label{eq:h2-BdG}
\end{equation}
with the total Hamiltonian $H^{\text{\tiny BdG}}_{\text{\tiny KW}} = \int\frac{d^3 {\bm p}}{(2\pi)^3} (\psi^{\text{\tiny BdG}}_{{\bm{p}}})^{\dagger}  h^{\text{\tiny BdG}}_{\text{\tiny KW}}({\bm{p}})  \psi^{\text{\tiny BdG}}_{{\bm{p}}}$.
Here $\tau_{1,2,3}$ are Pauli matrices acting on the fictitious Nambu-doubling index (particle at ${\bm{p}}$ vs.~hole at $-{\bm{p}}$). 
Consistency with the BdG redundancy or the particle-hole symmetry requires
$\tau_1\,h^{\text{\tiny BdG}}_{\text{\tiny KW}}({\bm{p}})^{*}\tau_1 = -\,h^{\text{\tiny BdG }}_{\text{\tiny KW}}(-{\bm{p}})$, 
which is indeed satisfied by~\eqref{eq:h2-BdG}.

What we have done so far is just rewriting the KW Hamiltonian from the standard representation to the BdG representation, thus it keeps all the properties including the symmetries.
For example, the $U(1)$ onsite symmetry is expressed in this representation as
\begin{equation}
[h^{\text{\tiny W}}_{\text{\tiny KW}}, \,\tau_{3}] \,=\, 0\,,
\end{equation}
which means
\begin{align}
&[H^{\text{\tiny BdG}}_{\text{\tiny KW}},\, Q_0] \,=\, 0\,,
\\
&Q_0 \,=\, \sum_{{\bm x}} (\psi^{\text{\tiny BdG}}_{\bm x})^{\dagger}\,\tau_3 \, \psi^{\text{\tiny BdG}}_{\bm x}
\,=\,  \int_{-\pi}^{\pi}\frac{d^3 p}{(2\pi)^3}\,
   (\psi^{\text{\tiny BdG}}_{{\bm{p}}})^{\dagger}\,\tau_3 \,
   \psi^{\text{\tiny BdG}}_{{\bm{p}}}\,.
\end{align}

To gap the Weyl node at ${\bm{p}}_B=(0,0,\pi)$ while keeping the one at ${\bm{p}}_A$ gapless, one adds a term
\begin{equation}
  \delta h(\bm{p})
  \,=\, \frac12\,\tau_1 \sigma_2(1-\cos p_3)\,,
  \label{eq:ssm1}
\end{equation}
which breaks the onsite $U(1)$ generated by $\tau_3$ in the Nambu space.
It is {\it a sort of species-splitting mass term, but not in the flavor space but in the Nambu (partice/hole) space.}
The resulting single-Weyl BdG Hamiltonian is
\begin{equation}
  h^{\text{\tiny BdG}}_{\text{\tiny single}}({\bm{p}})
  = \frac12\Bigl[
      \sigma_1 \sin p_1
      + \sigma_3 \sin p_2
        + \tau_3\sigma_2 \sin p_3
      + \sigma_2\, r \left( 2 - \cos p_1 - \cos p_2 \right)
      + \tau_1\sigma_2 (1-\cos p_3)
    \Bigr].
  \label{eq:h-single-BdG}
\end{equation}
For $|r|>1$ this Hamiltonian has a single Weyl node at ${\bm{p}}_A={\bm 0}$ and is fully gapped elsewhere in the Brillouin zone.
It has relatively small discrete symmetries; translational symmetry, 2d subgroup of cubic rotational symmetry, and BdG particle-hole redundancy.
The present single-Weyl Hamiltonian, however, has the exact non-onsite $U(1)$ symmetry as we will discuss later, which may have potential to prevent the generation of counterterms originating in the discrete symmetry breaking. This will be the main topic of Sec.~\ref{subsec:mu-deformation-and-tuning}.
For later convenience, we name the part of the Hamiltonian acting on the Nambu space nontrivially as
\begin{align}
N({\bm p})\,\equiv\,\frac{\sigma_2}{2}\,
\Bigl[\tau_3 \sin p_3  \,+\, \tau_1 (1-\cos p_3)\Bigr]
\,=\,\sigma_2\, \sin \frac{p_3}{2}\, \Bigl[\tau_3 \cos\frac{p_3}{2} \,+\, \tau_1 \sin \frac{p_3}{2}\Bigr]\,,
\label{eq:nontrivial}
\end{align}
which is interpreted as the effective one-dimensional Hamiltonian in the $p_3$ direction defined as $N({\bm p})\,\equiv\,h^{\text{\tiny BdG}}_{\text{\tiny single}}(p_1 = 0, p_2= 0, p_3)$.
We note that it satisfies $N({\bm p})^2 = (\sin \frac{p_3}{2})^2 {\bm 1}$.

To clarify how the Hamiltonian acts on the Nambu spinor $\psi^{\text{\tiny BdG}}_\p$, one can rewrite it as a matrix acting on the Nambu space,
\begin{equation}
h^{\text{\tiny BdG}}_{\text{\tiny single}}({\bm{p}}) = 
\frac{1}{2}
\begin{pmatrix} 
h^{\text{\tiny W}}_{\text{\tiny KW}}({\bm{p}}) & \Delta({\bm{p}}) \\ \Delta^\dagger({\bm{p}}) & -h^{\text{\tiny W}}_{\text{\tiny KW}}(-{\bm{p}})^{*} \end{pmatrix}\,,
\end{equation}
with the species-splitting mass $\Delta({\bm{p}})$
\begin{equation}
\Delta({\bm{p}}) = \sigma_2 (1-\cos p_3)  \,.
\end{equation}
This species-splitting mass term is a sort of Majorana mass term, giving the $\chi^{T}\chi$ term, who obviously breaks the $U(1)$ onsite symmetry.

\subsection{Non-onsite chiral symmetry and GW relation}
\label{eq:NOC}

The key observation of Ref.~\cite{Gioia:2025bhl} is that the single-Weyl
Hamiltonian~\eqref{eq:h-single-BdG} commutes with a momentum-dependent
generator
\begin{equation}
  S_{\chi}({\bm{p}})
  \,=\, \frac12\Bigl[\tau_3\,(1+\cos p_3)\, +\, \tau_1\,\sin p_3\Bigr]
  \,=\, 
  \cos\frac{p_3}{2} \,\Bigl[\tau_3 \cos\frac{p_3}{2} + \tau_1 \sin \frac{p_3}{2}\Bigr]\,.
  \label{eq:GT-Schiral}
\end{equation}
This generator is proportional to $N({\bm p})$, which is the only part acting on the Nambu space nontrivially in the Hamiltonian (\ref{eq:nontrivial}), thus it is obvious that it commutes with the Hamiltonian \eqref{eq:h-single-BdG}.
One can rephrase the situation as follows: The species-splitting mass term \eqref{eq:ssm1} rotates the $U(1)$ symmetric direction from $\tau_3$ to $\tau_3 \cos\frac{p_3}{2} + \tau_1 \sin \frac{p_3}{2}$ in the Nambu $SU(2)$ space, with gapping the one of the nodes.
Such a way of realization of chiral symmetry and doubler control on the lattice is found in the twisted-Wilson fermion \cite{Sint:2007ug, Shindler:2007vp} and the higher-Clifford-algebra fermion \cite{Misumi:2013maa}.

The ``chiral" symmetry is explicitly expressed as
\begin{equation}
  [S_{\chi}({\bm{p}}),\, h^{\text{\tiny BdG}}_{\text{\tiny single}}({\bm{p}})]\,=\,0
  \qquad\text{for all }{\bm{p}}\,,
\end{equation}
and $S_{\chi}$ obeys
\begin{equation}
 S_{\chi}({\bm{p}})^2
  = \left(\cos\frac{p_3}{2}\right)^2\,\bm{1}\,,
  \label{eq:GT-Schiral-square}
\end{equation}
which means that eigenvalues of this chiral generator is not quantized but has continuum spectrum.
Indeed, at the surviving Weyl node ${\bm{p}}_A = (0,0,0)$ one has
$S_{\chi}(\p_A)=\tau_3$, which coincides with the original charge operator in the BdG basis, while at the gapped node ${\bm{p}}_B = (0,0,\pi)$ one finds $S_{\chi}({\bm{p}}_B)=0$.

The associated many-body charge is
\begin{align}
Q_{\chi}
  \,=\, \int \frac{d^3{\bm p}}{(2\pi)^3}
    (\psi^{\text{\tiny BdG}}_{{\bm{p}}})^{\dagger}\,S_{\chi}({\bm{p}})\,\psi^{\text{\tiny BdG}}_{{\bm{p}}}
   \,=\, \frac12\sum_{{\bm{x}}}
    \Bigl(
      \chi^{\dagger}_{{\bm{x}}} \chi_{{\bm{x}}}
      + \chi^{\dagger}_{{\bm{x}}+{\bm e}_3}\chi_{\bm{x}}
      - i\,\chi^{\dagger}_{{\bm{x}}+{\bm e}_3}\chi^{\dagger}_{\bm{x}}
    \Bigr)
    + \text{h.c.}
    \,,
  \label{eq:GT-Qchiral-k}
\end{align}
which satisfies $[H^{\text{\tiny BdG}}_{\text{\tiny single}},\,Q_{\chi}]=0$.
This position-space expression involves onsite and nearest-neighbor hopping terms along the third axis.
In this sense the symmetry generated by $Q_{\chi}$ is manifestly non-onsite, yet it reduces to the usual axial $U(1)_A$ in the low-energy theory around $\bm{p}_A$.

The construction defines a consistent free field theory, but gauging it fails. The Majorana-like mass term $\chi^T \chi$ carries charge $2e$ under $U(1)$ gauge transformations, and this is the source where the gauge anomaly emerges in this formulation. As discussed in \cite{Gioia:2025bhl}, one needs to introduce flavors to cancel the gauge anomaly.

The chiral generator \eqref{eq:GT-Schiral} is interpreted as the Hamiltonian analogue of a Ginsparg--Wilson chiral operator: it is momentum-dependent, commutes with the Hamiltonian, and has a non-quantized spectrum constrained by~\eqref{eq:GT-Schiral-square}.
Indeed, for the effective one-dimensional Hamiltonian \eqref{eq:nontrivial} and the chiral generator \eqref{eq:GT-Schiral}, there is the nontrivial relation
\begin{align}
N({\bm p})^2 \,+\, S_{\chi}({\bm{p}})^2 \,=\, {\bm 1}\,.
\end{align}
It corresponds to the Hamiltoian version of Ginsparg--Wilson relation proposed in Ref.~\cite{Creutz:2001wp}, where it was already pointed out that the chiral generator for the Hamiltonian formalism needs to have the continuum spectrum.
There have been the proposal of the improved Ginsparg--Wilson relation to give the quantized spectrum \cite{Fujikawa:2000my, Singh:2025sye}.
It is an interesting avenue to consider the Hamiltonians satisfying such an improved relation.

\subsection{One-parameter deformation, extra nodes and need for tuning}
\label{subsec:mu-deformation-and-tuning}

In this subsection we discuss a natural one-parameter deformation of the single-Weyl BdG Hamiltonian and show that it can destabilize the ``single-Weyl'' situation by generating additional gapless nodes.
We then interpret this deformation as a radiatively-generated counterterm in an interacting theory and explain how a tuning of the parameter could be generically required.

Let us remind ourselves that the characteristic feature of the single-Weyl Hamiltonian \eqref{eq:h-single-BdG} is that all $\tau$-dependence sits in the single spin channel $\sigma_2$.
Therefore, a natural one-parameter deformation to the Hamiltonian \eqref{eq:h-single-BdG} is to add another $\sigma_2$-channel term with the same $\tau$ structure as $S_\chi(\p)$ in \eqref{eq:GT-Schiral}
\begin{equation}
\delta h_\mu(\bm p)
\,=\,
\frac{\mu}{2}\,\sigma_2\,\Big[\tau_3\,(1+\cos p_3)\,+\,\tau_1\,\sin p_3\,\Big]
\,=\,\mu\,\sigma_2\,S_{\chi}(\p).
\label{eq:Hmu_def}
\end{equation}
This deformation preserves the chiral symmetry as
\begin{equation}
[h^{\text{\tiny BdG}}_{\text{\tiny single}}({\bm{p}})+ \delta h_\mu(\bm p),\,S_{\chi}(\p)]\,=\,0
  \qquad\text{for all }{\bm{p}}\,,
\end{equation}
as well as all the other symmetries.
Physically, \eqref{eq:Hmu_def} shifts the Weyl node position $\p_A$ along the $p_3$ direction without introducing a conventional mass term.
We now write down the one-parameter-deformed Hamiltonian with collecting the $\tau$-dependent part in the $\sigma_2$ channel as
\begin{equation}
h(\bm p)\,\equiv\, h^{\text{\tiny BdG}}_{\text{\tiny single}}({\bm{p}})+ \delta h_\mu(\bm p)
\,=\,
\sigma_1\,\sin p_1\,+\,\sigma_3\,\sin p_2\,+\,\sigma_2\Big[rM(p_1,p_2)+a(p_3)\tau_3+b(p_3)\tau_1\Big],
\label{eq:H_total_r_mu}
\end{equation}
with
\begin{equation}
M(p_1,p_2)=2-\cos p_1 -\cos p_2 \,,
\quad
a(p_3)=\sin p_3+\mu(1+\cos p_3)\,,
\quad
b(p_3)=(1-\cos p_3)+\mu\sin p_3\,.
\label{eq:ab_def}
\end{equation}

Because $\sigma_1,\sigma_2,\sigma_3$ anticommute and the $\tau$ matrices commute with the $\sigma$ matrices, one finds
\begin{align}
4h(\bm p)^2
=
\sin^2 p_1+\sin^2 p_2+\Big[rM\,\mathbf 1_\tau+a\tau_3+b\tau_1\Big]^2\,,
\end{align}
Thus, the eigenvalues of $4h(\bm p)^2$ are
\begin{align}
\sin^2 p_1+\sin^2 p_2+\Big(rM\pm \sqrt{a^2+b^2}\Big)^2\,.
\end{align}
A necessary and sufficient condition for a gapless point is
\begin{equation}
\sin p_1=0,\quad \sin p_2=0,\quad rM(p_1,p_2)=\pm\sqrt{a(p_3)^2+b(p_3)^2}\,.
\label{eq:gapless_condition_master}
\end{equation}
Thus $p_1,p_2$ are restricted to the four values $(p_1,p_2)\in\{(0,0),\,(\pi,0),\,(0,\pi),\,(\pi,\pi)\}$.
At $(p_1,p_2)=(0,0)$ we have $M(p_1,p_2)=0$, hence the gapless condition reduces to $a(p_3)=0,\,b(p_3)=0$.
Using the parametrization $t=\tan(p_3/2)$ so that $\sin p_3=\frac{2t}{1+t^2}$ and $\cos p_3=\frac{1-t^2}{1+t^2}$, one finds
\begin{equation}
a(p_3)=0\quad\Longleftrightarrow\quad t=-\mu\,,
\qquad\text{and then}\qquad b(p_3)=0\,.
\end{equation}
Therefore the Weyl node is shifted from $p_3=0$ to
\begin{equation}
\tan\frac{p_3^\star}{2}=-\mu 
\qquad\Longleftrightarrow\qquad
 p_3^\star=-2\arctan\mu\quad (\mathrm{mod}\ 2\pi).
\label{eq:node_shift}
\end{equation}
Importantly, the would-be node at $p_3=\pi$ remains gapped because $b(p_3 = \pi)=2$ for any $\mu$.

The central issue is whether additional solutions of \eqref{eq:gapless_condition_master} appear at the other points in $(p_1,p_2)$.
To analyze this, we first simplify $\sqrt{a^2+b^2}$.
With $t=\tan(p_3/2)$, \eqref{eq:ab_def} becomes
$a(p_3)=\frac{2(t+\mu)}{1+t^2}\,,\,b(p_3)=\frac{2t(t+\mu)}{1+t^2}$,
hence
\begin{equation}
\sqrt{a(p_3)^2+b(p_3)^2}
\,=\,
2\sqrt{1+\mu^2}\,
\Big|\sin\Big(\frac{p_3}{2}+\phi\Big)\Big|\,,
\qquad
\phi\equiv \arctan\mu\,.
\label{eq:ab_norm_sine_form}
\end{equation}
In particular,
\begin{equation}
0\le \sqrt{a^2+b^2}\le 2\sqrt{1+\mu^2}.
\label{eq:ab_max}
\end{equation}

The first possible extra nodes are $(p_1,p_2)=(\pi,0)$ and $(0,\pi)$.
For these points we have $M(p_1,p_2)=2$, so the gapless condition \eqref{eq:gapless_condition_master} reads
\begin{equation}
2r=\sqrt{a(p_3)^2+b(p_3)^2}.
\label{eq:extra_TRIM_pi0}
\end{equation}
By \eqref{eq:ab_max}, \eqref{eq:extra_TRIM_pi0} has a solution in $p_3$ if and only if
\begin{equation}
2r\le 2\sqrt{1+\mu^2}
\qquad\Longleftrightarrow\qquad
r\le \sqrt{1+\mu^2}
\qquad\Longleftrightarrow\qquad
|\mu|\ge \sqrt{r^2-1} \quad (|r|>1).
\label{eq:mu_threshold_main}
\end{equation}
For generic parameters with strict inequality, $|\mu|>\sqrt{r^2-1}$, there are in fact two solutions for $p_3$ at each of $(\pi,0)$ and $(0,\pi)$, because \eqref{eq:ab_norm_sine_form} gives
\begin{equation}
\Big|\sin\Big(\frac{\hat p_3}{2}+\phi\Big)\Big|=\frac{r}{\sqrt{1+\mu^2}}
\qquad\Rightarrow\qquad
\hat p_3=-2\phi \pm 2\arcsin\!\Big(\frac{r}{\sqrt{1+\mu^2}}\Big)\quad (\mathrm{mod}\ 2\pi).
\label{eq:kz_extra_nodes_explicit}
\end{equation}
Thus, once the bound $|\mu|<\sqrt{r^2-1}$ is violated, the system is no longer in the phase with a single Weyl node: additional gapless nodes are created at $(\pi,0,\hat p_3)$ and $(0,\pi,\hat p_3)$. 

Further possible nodes can be at $(p_1,p_2)=(\pi,\pi)$.
At $(\pi,\pi)$ we have $M(p_1,p_2)=4$, and the condition becomes $4r=\sqrt{a^2+b^2}$, which can be satisfied only for even larger $|\mu|$:
\begin{equation}
4r\le 2\sqrt{1+\mu^2}
\qquad\Longleftrightarrow\qquad
|\mu|\ge \sqrt{4r^2-1}\,.
\end{equation}
Hence the first instability of the single-Weyl regime is governed by \eqref{eq:mu_threshold_main}.

In summary, the $\mu$-deformation \eqref{eq:Hmu_def} shifts the physical Weyl node as in \eqref{eq:node_shift}, but it also opens a channel for new gapless nodes to appear once
\begin{equation}
|\mu|\geq\sqrt{r^2-1}\,.
\end{equation}
It is notable that the number of species becomes five for $\sqrt{r^2-1}<|\mu|<\sqrt{4r^2-1}$ since $\hat p_3$ are two-folded.
It is notable that the Karsten-Wilczek Dirac operator has six species for a certain parameter domain \cite{Misumi:2012uu} and one of the species is now gapped due to the species-splitting mass.

The deformation \eqref{eq:Hmu_def} is not an arbitrary modification: it is a local, symmetry-allowed operator in the same $\sigma_2$ channel as the original $\tau$ structure, and it commutes with the chiral generator $S_\chi(\p)$.
In an interacting gauge theory, any local operator not forbidden by symmetries is expected to be generated by radiative corrections.
From this viewpoint, $\mu$ should be treated as a counterterm coefficient
\begin{equation}
\mu_{\rm R}=\mu_0+\delta\mu(g^2,r,a)\,,
\end{equation}
where $\mu_0$ is the bare parameter in the lattice action and $\delta\mu$ is the radiative correction.
By the same logic as the additive mass renormalization in Wilson fermions, $\delta\mu$ is not parametrically protected and may be $O(1)$ in lattice units unless an additional symmetry suppresses it. 

It is notable that the counterterm \eqref{eq:Hmu_def} explicitly breaks the cubic rotational symmetry, P and T symmetries, which are already broken in the undeformed Hamiltonian. Thus, we can interpret that the radiative correction \eqref{eq:Hmu_def} in the interacting theory originates in the breaking of those symmetries.
It is a similar situation to the necessity of tuning the dimension-3 and -4 counterterms in the lattice gauge simulation with the Lagrangian minimal-doubling formalism \cite{Bedaque:2008xs,Bedaque:2008jm, Capitani:2009yn,Capitani:2010nn,Capitani:2013zta,Capitani:2013iha,Weber:2013tfa,Weber:2017eds}.

The analysis above shows that the ``single-Weyl'' regime is not guaranteed merely by enforcing the chiral symmetry $S_\chi(\p)$: it requires the renormalized parameter to remain within the single-Weyl domain,
\begin{equation}
|\mu_{\rm R}|<\sqrt{r^2-1}\,.
\label{eq:single_weyl_domain}
\end{equation}
Therefore, to maintain a single Weyl node in the interacting theory, one must in general
\begin{itemize}
\item include the $\mu$--term \eqref{eq:Hmu_def} from the outset as an allowed counterterm, and
\item adjust $\mu_0$ (as a function of the gauge coupling and lattice spacing) so that \eqref{eq:single_weyl_domain} holds,
\end{itemize}
unless an extra symmetry is imposed that forbids \eqref{eq:Hmu_def} altogether.

As we have discussed, the single-Weyl Hamiltonian itself cannot directly be gauged because of the gauge anomaly and the gauging requires introduction of flavor degrees of freedom \cite{Gioia:2025bhl}. We emphasize, however, that the emergence of extra nodes in interacting theories can occur broadly in multi-flavor anomaly-free setups and other exactly chiral-symmetric Weyl formulations.   

In the end of the section, we make a comment on the recent work \cite{Meyniel:2025euu}. In the work, the authors in detail argued the classification of the single-Weyl Hamiltonians including the proposal of Gioia-Thorngren \cite{Gioia:2025bhl}. They studied the change of the node configuration by varying parameters corresponding to a magnetic field ($r$ in our notation) and mass or chemical potential, but did not consider the symmetry-preserving ``chiral-generator" deformation we propose here. It is an interesting question how the node configuration is affected if these independent deformations are combined in the variety of single-Weyl Hamiltonians.


\section{Conclusion and discussion}
\label{sec:CD}

In this work, we developed a systematic Hamiltonian framework for minimally doubled lattice fermions in three spatial dimensions. We treated both four-component Dirac and two-component Weyl constructions, and classified the resulting Hamiltonians by their nodal structure and discrete symmetry patterns. For representative examples, we derived explicit Bloch Hamiltonians, determined the node locations, and identified the discrete symmetries that remain intact and those that are broken.

We then reconsidered BdG-based ``single-Weyl'' Hamiltonians within our minimally doubled framework. By introducing a Nambu particle/hole doubling and an appropriate species-splitting ``Majorana" mass term into the Karsten-Wilczek minimal-doubling Hamiltonian, we showed that one Weyl node can be gapped out while the other remains gapless in the free theory. The protection mechanism is encoded in a momentum-dependent conserved generator acting in Nambu space: it commutes with the Hamiltonian, yet its spectrum is not quantized, which naturally suggests the Hamiltonian Ginsparg-Wilson-type structure. From this viewpoint, we studied a symmetry-preserving one-parameter deformation that keeps the full symmetry intact, yet can destabilize the single-node regime by generating additional Weyl nodes once the deformation exceeds a critical threshold. Concretely, the deformation opens a channel for extra gapless solutions that were gapped in the undeformed model, and the single-Weyl regime survives only inside a finite parameter window.

An important implication for the interacting theories is that the destabilizing deformation is not an artificial modification: it is not forbidden by the exact non-onsite conservation law, and accordingly, once gauge interactions are introduced, radiative corrections are expected to generate this term as an allowed counterterm with a renormalized coefficient that is not parametrically suppressed in lattice units, in close analogy with additive mass renormalization in Wilson fermion. The stability of the single-Weyl regime therefore could require a moderate tuning of the corresponding bare parameter so that the renormalized coefficient remains inside the single-node domain. In this sense, the exact non-onsite conservation alone does not guarantee radiative stability of the single-Weyl phase.

Finally, let us comment on the relation of the present constructions to the Nielsen-Ninomiya no-go theorem. Under its standard assumptions, including locality, Hermiticity, translation invariance, and conserved onsite charges with discrete spectra, an isolated Weyl mode is forbidden. In the minimal-doubling constructions, the four-component Hamiltonian, which possesses two conserved onsite charges, yields two massless Dirac fermions, while the two-component Hamiltonian, with one conserved onsite charge, yields one massless Dirac fermion composed of two Weyl fermions. These constructions therefore remain within the standard scope of the theorem. By contrast, the BdG-based single-Weyl Hamiltonian possesses only a non-onsite momentum-dependent conserved generator, and hence lies outside that scope.

Several directions are suggested by our analysis. 
First, it is natural to ask whether one can engineer single-node BdG constructions starting from minimal-doubling Hamiltonians beyond the Karsten-Wilczek prototype, in such a way that the protecting symmetry is strengthened or that the discrete rotational structure is improved. From the minimally doubled viewpoint, this becomes a concrete design problem: identify which minimal-doubling classes admit a Nambu-space ``species-splitting'' mechanism while preserving the largest possible remnant lattice symmetry.
Second, our results call for explicit numerical studies in the interacting theory. Given a candidate single-Weyl Hamiltonian (with or without deformations), one can directly verify the single-node condition by scanning the Brillouin zone for gapless points and by monitoring how the node count changes as the parameter vary. Such computations would provide a sharp, nonperturbative test of the analytic stability bounds and would clarify how robust the single-Weyl regime remains once interactions are included.
Third, it is important to map out the phase structure in the enlarged parameter space, in the spirit of earlier analyses of minimally doubled fermions in Lagrangian formulations \cite{Misumi:2012uu}. A systematic ``phase diagram'' of node configurations, including the critical surfaces where extra nodes are created or annihilated, would both quantify the required tuning and help identify parameter regions where radiative stability might be enhanced by additional accidental symmetries or by improved lattice point-group structure.

From the condensed-matter viewpoint, single-Weyl Hamiltonians may serve as useful lattice models of Weyl-semimetal- or gapless-superconductor-type systems for studying how an isolated Weyl node can appear, how it can be destabilized by symmetry-allowed perturbations, and how additional nodes can emerge as parameters are varied.

\begin{acknowledgments} 
The author appreciates the YITP workshop ``Progress and Future of non-perturbative quantum field theory" (YITP-W-25-25) and the YITP long-term workshop ``Hadrons and Hadron Interactions in QCD 2024'' (YITP-T-24-02) for providing the opportunities of useful discussions.  This work was partially supported by Japan Society for the Promotion of Science (JSPS) KAKENHI Grant No.~23K03425 and 22H05118.
\end{acknowledgments}


\appendix

\section{Minimal-doubling Dirac operators}
\label{sec:KW-CKM-review}

In this appendix, based on Ref.~\cite{Creutz:2010qm, Creutz:2010bm}, we review a four-dimensional Euclidean lattice formulation of minimally-doubled fermions, with focus on the point-split flavor fields and the flavored-mass terms.
For recent numerical and analytical developments with the point-split fields of minimal-doubling fermions, see Refs.~\cite{Shukre:2025bnm, Weber:2025kcl, Shukre:2024tkw, Weber:2023kth, Borsanyi:2025big, Vig:2024umj, Durr:2024ttb}.
We here set the lattice spacing to unity and consider a four-component Dirac spinor \(\psi(x)\). A Karsten-Wilczek minimally doubled Dirac operator in momentum space is
\begin{equation}
  D_{\rm KW}(p)
  = i\sum_{j=1}^4 \gamma_j \sin p_j
    + i\gamma_4 (3 - \cos p_1 - \cos p_2 - \cos p_3)\,.
  \label{eq:KW-Dirac-0pi}
\end{equation}
One easily checks that
\begin{equation}
  D_{\rm KW}(p) = 0
  \quad\Longleftrightarrow\quad
  p_A = (0,0,0,0)\,\,\, \text{or}\,\,\, p_B = (0,0,0,\pi)\,,
  \label{eq:KW-poles-0pi}
\end{equation}
while all other would-be doublers are lifted by the Wilson-like term.

At the two zeros the Dirac structure is inequivalent. The gamma matrices at \(p_A\) are related to those at \(p_B\) by a similarity transformation
\begin{equation}
  \gamma_\mu ' = \Gamma^{-1}\gamma_\mu\Gamma\,,\qquad
  \Gamma = i\gamma_4\gamma_5 \,.
  \label{eq:Gamma-def}
\end{equation}
This difference will be taken into account through a point-splitting procedure that identifies the two poles as independent flavors.

To make the two minimally doubled species manifest, we introduce momentum-dependent factors that project onto the neighborhoods of \(p_A\) and \(p_B\). Since in our convention the poles lie at these two values, it is natural to use
\begin{equation}
  P_A = \frac{1+\cos p_4}{2}\,,\qquad
  P_B = \frac{1-\cos p_4}{2}\,.
  \label{eq:P12-0pi-def}
\end{equation}
Near \(p_A\) one has \(\cos p_4\simeq 1\) and hence $P_A\simeq 1,\, P_B\simeq 0 \quad (p_4\simeq 0)$, whereas near \(p_B\) one has \(\cos p_4\simeq -1\), so that $P_A\simeq 0,\, P_B\simeq 1 \quad (p_4\simeq \pi)$. We then define two point-split Dirac fields in momentum space by
\begin{equation}
  u(p-p_A) = P_A\,\psi(p)\,,\qquad
  d(p-p_B) = \Gamma\,P_B\,\psi(p)\,,
  \label{eq:ud-0pi-def}
\end{equation}
where \(\Gamma\) is the matrix in \eqref{eq:Gamma-def}. By construction, \(u(p-p_A)\) has support near the pole at \(p_4=0\), whereas \(d(p-p_B)\) has support near \(p_4=\pi\). In position space, these fields correspond to linear combinations of \(\psi(x)\) and its nearest neighbors in the 4-direction.

We now assemble the two point-split fields into a flavor doublet
\begin{equation}
  \Psi_{\rm f}(p)
  =
  \begin{pmatrix}
    u(p-p_A)\\[0.2em]
    d(p-p_B)
  \end{pmatrix}\,,
  \label{eq:Psi-flavor-doublet-0pi}
\end{equation}
on which Pauli matrices \(\tau_i\) act in flavor (species) space. Because of the factor \(\Gamma\) in \eqref{eq:ud-0pi-def}, the Dirac structure at the two poles is brought to a common representation, and the action of the usual axial generator \(\gamma_5\) becomes
\begin{equation}
  \gamma_5\,\psi(p)
  \ \longrightarrow\
  \begin{pmatrix}
    +\gamma_5 & 0\\
    0 & -\gamma_5
  \end{pmatrix}
  \Psi_{\rm f}(p)
  = (\gamma_5\otimes\tau_3)\,\Psi_{\rm f}(p)\,.
  \label{eq:gamma5-flavored-0pi}
\end{equation}
That is, in the two-flavor description the exact lattice chiral symmetry of the KW action is realized as a flavored U(1) generated by
\(\gamma_5\otimes\tau_3\), rather than by the naive singlet
\(\gamma_5\otimes\bm{1}\).

The point-split fields \(u\) and \(d\) allow us to define flavored mass
terms that assign different masses to the two species.  In the flavor basis the simplest such term is
\begin{equation}
m\int \frac{d^4 p}{(2\pi)^4} \bar\Psi_{\rm f}(p)\,
      \bigl(\bm{1}\otimes\tau_3\bigr)\,
      \Psi_{\rm f}(p)
  = m\int \frac{d^4 p}{(2\pi)^4} \bigl[\bar u(p-p_A)u(p-p_A)-\bar d(p-p_B)d(p-p_B)\bigr]\,,
  \label{eq:flavored-mass-flavor-basis}
\end{equation}
with a real parameter \(m\).
In terms of the original field \(\psi(p)\), this becomes a momentum-dependent ``flavored'' mass.
Using the definitions in \eqref{eq:ud-0pi-def} and the fact that
\(\Gamma^\dagger\Gamma=\bm{1}\), one finds
$ \bar u(p-p_A)u(p-p_A) = \bar\psi(p) P_A^2 \psi(p),\,\, \bar d(p-p_B)d(p-p_B) = \bar\psi(p) P_B^2 \psi(p)$,
so that
\begin{equation}
  \bar u(p-p_A)u(p-p_A)-\bar d(p-p_B)d(p-p_B)
  = \bigl[P_A^2 - P_B^2\bigr]\bar\psi(p)\psi(p)
  = \cos p_4 \bar\psi(p)\psi(p)\,,
\end{equation}
and hence the flavored mass term in the original Dirac field representation, which keeps the mode at $p_A$ massless and makes the other mode at $p_B$ massive, reads
\begin{equation}
  S_{\rm m}^{\rm flavor}
  = m\int \frac{d^4 p}{(2\pi)^4}(1- \cos p_4)\,\bar\psi(p)\psi(p)\,.
  \label{eq:flavored-mass-cosp4}
\end{equation}
Near the two poles this behaves as $\cos p_4 \sim + 1$ $(p_4\simeq 0)$ and
$\cos p_4 \sim -1$ $(p_4\simeq \pi)$,
so that one of the species acquires bare mass \(2m\) while the other remains massless.
In position space the factor \(\cos p_4\) corresponds to a symmetric
nearest-neighbor hopping in the 4-direction.
Introducing the translation operators
 $ (T_{\pm4}\psi)(x) = U_{x,\pm4}\,\psi({x\pm e_4})$,
one obtains
\begin{equation}
  S_{\rm m}^{\rm flavor}
  = m\sum_x \bar\psi(x)\,[1-\bigl(T_{+4}+T_{-4}\bigr)/2]\,\psi(x) \,.
  \label{eq:flavored-mass-position-0pi}
\end{equation}
This is a local, gauge-covariant operator involving only nearest neighbors along the temporal direction.



\bibliographystyle{utphys}
\bibliography{./QFT,./refs}

\providecommand{\href}[2]{#2}\begingroup\raggedright\begin{thebibliography}{10}

\bibitem{Wilson:1974sk}
K.~G. Wilson, ``{Confinement of Quarks},''
\href{http://dx.doi.org/10.1103/PhysRevD.10.2445}{{\em Phys. Rev.} {\bfseries
  D10} (1974) 2445--2459}.

\bibitem{Creutz:1980zw}
M.~Creutz, ``{Monte Carlo Study of Quantized SU(2) Gauge Theory},''
\href{http://dx.doi.org/10.1103/PhysRevD.21.2308}{{\em Phys. Rev.} {\bfseries
  D21} (1980) 2308--2315}.

\bibitem{Karsten:1980wd}
L.~H. Karsten and J.~Smit, ``{Lattice Fermions: Species Doubling, Chiral
  Invariance, and the Triangle Anomaly},''
  \href{http://dx.doi.org/10.1016/0550-3213(81)90549-6}{{\em Nucl. Phys. B}
  {\bfseries 183} (1981) 103}.

\bibitem{Nielsen:1980rz}
H.~B. Nielsen and M.~Ninomiya, ``{Absence of Neutrinos on a Lattice. 1. Proof
  by Homotopy Theory},''
\href{http://dx.doi.org/10.1016/0550-3213(81)90361-8}{{\em Nucl. Phys.}
  {\bfseries B185} (1981) 20}.

\bibitem{Nielsen:1981xu}
H.~B. Nielsen and M.~Ninomiya, ``{Absence of Neutrinos on a Lattice. 2.
  Intuitive Topological Proof},''
\href{http://dx.doi.org/10.1016/0550-3213(81)90524-1}{{\em Nucl. Phys.}
  {\bfseries B193} (1981) 173--194}.

\bibitem{Nielsen:1981hk}
H.~B. Nielsen and M.~Ninomiya, ``{No Go Theorem for Regularizing Chiral
  Fermions},''
\href{http://dx.doi.org/10.1016/0370-2693(81)91026-1}{{\em Phys. Lett.}
  {\bfseries 105B} (1981) 219--223}.

\bibitem{Wilson:1975id}
K.~G. Wilson, \href{http://dx.doi.org/10.1007/978-1-4613-4208-3_6}{``{Quarks
  and Strings on a Lattice},''} in {\em {New Phenomena in Subnuclear Physics:
  Proceedings, International School of Subnuclear Physics, Erice, Sicily, Jul
  11-Aug 1 1975. Part A}}, pp.~69--142.
\newblock 1975.
\newblock
\url{https://doi.org/10.1007/978-1-4613-4208-3_6}.
\newblock

\bibitem{Kogut:1974ag}
J.~B. Kogut and L.~Susskind, ``{Hamiltonian Formulation of Wilson's Lattice
  Gauge Theories},''
\href{http://dx.doi.org/10.1103/PhysRevD.11.395}{{\em Phys. Rev.} {\bfseries
  D11} (1975) 395--408}.

\bibitem{Susskind:1976jm}
L.~Susskind, ``{Lattice Fermions},''
\href{http://dx.doi.org/10.1103/PhysRevD.16.3031}{{\em Phys. Rev.} {\bfseries
  D16} (1977) 3031--3039}.

\bibitem{Kawamoto:1981hw}
N.~Kawamoto and J.~Smit, ``{Effective Lagrangian and Dynamical Symmetry
  Breaking in Strongly Coupled Lattice QCD},''
  \href{http://dx.doi.org/https://doi.org/10.1016/0550-3213(81)90196-6}{{\em
  Nucl. Phys. B} {\bfseries 192} no.~1, (1981) 100--124}.
  \url{https://www.sciencedirect.com/science/article/pii/0550321381901966}.

\bibitem{Sharatchandra:1981si}
H.~S. Sharatchandra, H.~J. Thun, and P.~Weisz, ``{Susskind Fermions on a
  Euclidean Lattice},''
\href{http://dx.doi.org/10.1016/0550-3213(81)90200-5}{{\em Nucl. Phys.}
  {\bfseries B192} (1981) 205--236}.

\bibitem{Golterman:1984cy}
M.~F. Golterman and J.~Smit, ``{Selfenergy and Flavor Interpretation of
  Staggered Fermions},''
  \href{http://dx.doi.org/10.1016/0550-3213(84)90424-3}{{\em Nucl. Phys. B}
  {\bfseries 245} (1984) 61--88}.

\bibitem{Golterman:1985dz}
M.~F. Golterman, ``{STAGGERED MESONS},''
  \href{http://dx.doi.org/10.1016/0550-3213(86)90383-4}{{\em Nucl. Phys. B}
  {\bfseries 273} (1986) 663--676}.

\bibitem{Kilcup:1986dg}
G.~Kilcup and S.~R. Sharpe, ``{A Tool Kit for Staggered Fermions},''
  \href{http://dx.doi.org/10.1016/0550-3213(87)90285-9}{{\em Nucl. Phys. B}
  {\bfseries 283} (1987) 493--550}.

\bibitem{Kaplan:1992bt}
D.~B. Kaplan, ``{A Method for simulating chiral fermions on the lattice},''
  \href{http://dx.doi.org/10.1016/0370-2693(92)91112-M}{{\em Phys. Lett.}
  {\bfseries B288} (1992) 342--347},
\href{http://arxiv.org/abs/hep-lat/9206013}{{\ttfamily arXiv:hep-lat/9206013
  [hep-lat]}}.

\bibitem{Shamir:1993zy}
Y.~Shamir, ``{Chiral fermions from lattice boundaries},''
  \href{http://dx.doi.org/10.1016/0550-3213(93)90162-I}{{\em Nucl. Phys.}
  {\bfseries B406} (1993) 90--106},
\href{http://arxiv.org/abs/hep-lat/9303005}{{\ttfamily arXiv:hep-lat/9303005
  [hep-lat]}}.

\bibitem{Furman:1994ky}
V.~Furman and Y.~Shamir, ``{Axial symmetries in lattice QCD with Kaplan
  fermions},'' \href{http://dx.doi.org/10.1016/0550-3213(95)00031-M}{{\em Nucl.
  Phys. B} {\bfseries 439} (1995) 54--78},
  \href{http://arxiv.org/abs/hep-lat/9405004}{{\ttfamily
  arXiv:hep-lat/9405004}}.

\bibitem{Neuberger:1998wv}
H.~Neuberger, ``{More about exactly massless quarks on the lattice},''
  \href{http://dx.doi.org/10.1016/S0370-2693(98)00355-4}{{\em Phys. Lett.}
  {\bfseries B427} (1998) 353--355},
\href{http://arxiv.org/abs/hep-lat/9801031}{{\ttfamily arXiv:hep-lat/9801031
  [hep-lat]}}.

\bibitem{Ginsparg:1981bj}
P.~H. Ginsparg and K.~G. Wilson, ``{A Remnant of Chiral Symmetry on the
  Lattice},''
\href{http://dx.doi.org/10.1103/PhysRevD.25.2649}{{\em Phys. Rev.} {\bfseries
  D25} (1982) 2649}.

\bibitem{Luscher:1998pqa}
M.~Luscher, ``{Exact chiral symmetry on the lattice and the Ginsparg-Wilson
  relation},'' \href{http://dx.doi.org/10.1016/S0370-2693(98)00423-7}{{\em
  Phys. Lett. B} {\bfseries 428} (1998) 342--345},
  \href{http://arxiv.org/abs/hep-lat/9802011}{{\ttfamily
  arXiv:hep-lat/9802011}}.

\bibitem{Bietenholz:1999km}
W.~Bietenholz and I.~Hip, ``{The Scaling of exact and approximate
  Ginsparg-Wilson fermions},''
  \href{http://dx.doi.org/10.1016/S0550-3213(99)00477-0}{{\em Nucl. Phys. B}
  {\bfseries 570} (2000) 423--451},
  \href{http://arxiv.org/abs/hep-lat/9902019}{{\ttfamily
  arXiv:hep-lat/9902019}}.

\bibitem{Creutz:2010bm}
M.~Creutz, T.~Kimura, and T.~Misumi, ``{Index Theorem and Overlap Formalism
  with Naive and Minimally Doubled Fermions},''
  \href{http://dx.doi.org/10.1007/JHEP12(2010)041}{{\em JHEP} {\bfseries 12}
  (2010) 041},
\href{http://arxiv.org/abs/1011.0761}{{\ttfamily arXiv:1011.0761 [hep-lat]}}.

\bibitem{Durr:2010ch}
S.~D\"urr and G.~Koutsou, ``{Brillouin improvement for Wilson fermions},''
  \href{http://dx.doi.org/10.1103/PhysRevD.83.114512}{{\em Phys. Rev. D}
  {\bfseries 83} (Jun, 2011) 114512},
  \href{http://arxiv.org/abs/1012.3615}{{\ttfamily arXiv:1012.3615 [hep-lat]}}.
  \url{https://link.aps.org/doi/10.1103/PhysRevD.83.114512}.

\bibitem{Durr:2012dw}
S.~D\"urr, G.~Koutsou, and T.~Lippert, ``{Meson and Baryon dispersion relations
  with Brillouin fermions},''
  \href{http://dx.doi.org/10.1103/PhysRevD.86.114514}{{\em Phys. Rev. D}
  {\bfseries 86} (2012) 114514},
  \href{http://arxiv.org/abs/1208.6270}{{\ttfamily arXiv:1208.6270 [hep-lat]}}.

\bibitem{Misumi:2012eh}
T.~Misumi, ``{New fermion discretizations and their applications},''
  \href{http://dx.doi.org/10.22323/1.164.0005}{{\em PoS} {\bfseries
  LATTICE2012} (2012) 005},
\href{http://arxiv.org/abs/1211.6999}{{\ttfamily arXiv:1211.6999 [hep-lat]}}.

\bibitem{Cho:2013yha}
Y.-G. Cho, S.~Hashimoto, J.-I. Noaki, A.~Juttner, and M.~Marinkovic,
  ``{$O(a^2)$-improved actions for heavy quarks and scaling studies on quenched
  lattices},'' \href{http://dx.doi.org/10.22323/1.187.0255}{{\em PoS}
  {\bfseries LATTICE2013} (2014) 255},
  \href{http://arxiv.org/abs/1312.4630}{{\ttfamily arXiv:1312.4630 [hep-lat]}}.

\bibitem{Cho:2015ffa}
Y.-G. Cho, S.~Hashimoto, A.~Juttner, T.~Kaneko, M.~Marinkovic, J.-I. Noaki, and
  J.~T. Tsang, ``{Improved lattice fermion action for heavy quarks},''
  \href{http://dx.doi.org/10.1007/JHEP05(2015)072}{{\em JHEP} {\bfseries 05}
  (2015) 072}, \href{http://arxiv.org/abs/1504.01630}{{\ttfamily
  arXiv:1504.01630 [hep-lat]}}.

\bibitem{Durr:2017wfi}
S.~D\"urr and G.~Koutsou, ``{On the suitability of the Brillouin action as a
  kernel to the overlap procedure},''
  \href{http://arxiv.org/abs/1701.00726}{{\ttfamily arXiv:1701.00726
  [hep-lat]}}.

\bibitem{Adams:2009eb}
D.~H. Adams, ``Theoretical foundation for the index theorem on the lattice with
  staggered fermions,''
  \href{http://dx.doi.org/10.1103/PhysRevLett.104.141602}{{\em Phys.Rev.Lett.}
  {\bfseries 104} (2010) 141602},
  \href{http://arxiv.org/abs/0912.2850}{{\ttfamily arXiv:0912.2850 [hep-lat]}}.

\bibitem{Adams:2010gx}
D.~H. Adams, ``Pairs of chiral quarks on the lattice from staggered fermions,''
  \href{http://dx.doi.org/10.1016/j.physletb.2011.04.034}{{\em Phys.Lett.B}
  {\bfseries 699} (2011) 394--397},
  \href{http://arxiv.org/abs/1008.2833}{{\ttfamily arXiv:1008.2833 [hep-lat]}}.

\bibitem{Hoelbling:2010jw}
C.~Hoelbling, ``Single flavor staggered fermions,''
  \href{http://dx.doi.org/10.1016/j.physletb.2010.12.062}{{\em Phys.Lett.B}
  {\bfseries 696} (2011) 422--425},
  \href{http://arxiv.org/abs/1009.5362}{{\ttfamily arXiv:1009.5362 [hep-lat]}}.

\bibitem{deForcrand:2011ak}
P.~de~Forcrand, A.~Kurkela, and M.~Panero, ``{Numerical properties of staggered
  overlap fermions},'' {\em PoS} {\bfseries LATTICE2010} (2010) 080,
  \href{http://arxiv.org/abs/1102.1000}{{\ttfamily arXiv:1102.1000 [hep-lat]}}.

\bibitem{Creutz:2011cd}
M.~Creutz, T.~Kimura, and T.~Misumi, ``{Aoki Phases in the Lattice Gross-Neveu
  Model with Flavored Mass terms},''
  \href{http://dx.doi.org/10.1103/PhysRevD.83.094506}{{\em Phys. Rev.}
  {\bfseries D83} (2011) 094506},
\href{http://arxiv.org/abs/1101.4239}{{\ttfamily arXiv:1101.4239 [hep-lat]}}.

\bibitem{Misumi:2011su}
T.~Misumi, M.~Creutz, T.~Kimura, T.~Z. Nakano, and A.~Ohnishi, ``{Aoki Phases
  in Staggered-Wilson Fermions},''
  \href{http://dx.doi.org/10.22323/1.139.0108}{{\em PoS} {\bfseries
  LATTICE2011} (2011) 108}, \href{http://arxiv.org/abs/1110.1231}{{\ttfamily
  arXiv:1110.1231 [hep-lat]}}.

\bibitem{Follana:2011kh}
E.~Follana, V.~Azcoiti, G.~Di~Carlo, and A.~Vaquero, ``{Spectral Flow and Index
  Theorem for Staggered Fermions},''
  \href{http://dx.doi.org/10.22323/1.139.0100}{{\em PoS} {\bfseries
  LATTICE2011} (2011) 100}, \href{http://arxiv.org/abs/1111.3502}{{\ttfamily
  arXiv:1111.3502 [hep-lat]}}.

\bibitem{deForcrand:2012bm}
P.~de~Forcrand, A.~Kurkela, and M.~Panero, ``Numerical properties of staggered
  quarks with a taste-dependent mass term,''
  \href{http://dx.doi.org/10.1007/JHEP04(2012)142}{{\em JHEP} {\bfseries 04}
  (2012) 142}, \href{http://arxiv.org/abs/1202.1867}{{\ttfamily arXiv:1202.1867
  [hep-lat]}}.

\bibitem{Misumi:2012sp}
T.~Misumi, T.~Z. Nakano, T.~Kimura, and A.~Ohnishi, ``Strong-coupling analysis
  of parity phase structure in staggered-wilson fermions,''
  \href{http://dx.doi.org/10.1103/PhysRevD.86.034501}{{\em Phys.Rev.D}
  {\bfseries 86} (2012) 034501},
  \href{http://arxiv.org/abs/1205.6545}{{\ttfamily arXiv:1205.6545 [hep-lat]}}.

\bibitem{Durr:2013gp}
S.~D\"urr, ``{Taste-split staggered actions: eigenvalues, chiralities and
  Symanzik improvement},''
  \href{http://dx.doi.org/10.1103/PhysRevD.87.114501}{{\em Phys. Rev. D}
  {\bfseries 87} (Jun, 2013) 114501},
  \href{http://arxiv.org/abs/1302.0773}{{\ttfamily arXiv:1302.0773 [hep-lat]}}.
  \url{https://link.aps.org/doi/10.1103/PhysRevD.87.114501}.

\bibitem{Hoelbling:2016qfv}
C.~Hoelbling and C.~Zielinski, ``{Spectral properties and chiral symmetry
  violations of (staggered) domain wall fermions in the Schwinger model},''
  \href{http://dx.doi.org/10.1103/PhysRevD.94.014501}{{\em Phys. Rev. D}
  {\bfseries 94} no.~1, (2016) 014501},
  \href{http://arxiv.org/abs/1602.08432}{{\ttfamily arXiv:1602.08432
  [hep-lat]}}.

\bibitem{Zielinski:2017pko}
C.~Zielinski, {\em {Theoretical and Computational Aspects of New Lattice
  Fermion Formulations}}.
\newblock PhD thesis, Nanyang Technol. U., 2016.
\newblock \href{http://arxiv.org/abs/1703.06364}{{\ttfamily arXiv:1703.06364
  [hep-lat]}}.

\bibitem{Kimura:2011ik}
T.~Kimura, S.~Komatsu, T.~Misumi, T.~Noumi, S.~Torii, and S.~Aoki,
  ``{Revisiting symmetries of lattice fermions via spin-flavor
  representation},'' \href{http://dx.doi.org/10.1007/JHEP01(2012)048}{{\em
  JHEP} {\bfseries 01} (2012) 048},
\href{http://arxiv.org/abs/1111.0402}{{\ttfamily arXiv:1111.0402 [hep-lat]}}.

\bibitem{Chowdhury:2013ux}
A.~Chowdhury, A.~Harindranath, J.~Maiti, and S.~Mondal, ``Many avatars of the
  wilson fermion: A perturbative analysis,''
  \href{http://dx.doi.org/10.1007/JHEP02(2013)037}{{\em JHEP} {\bfseries 02}
  (2013) 037}, \href{http://arxiv.org/abs/1301.0675}{{\ttfamily arXiv:1301.0675
  [hep-lat]}}.

\bibitem{Misumi:2019jrt}
T.~Misumi and Y.~Tanizaki, ``{Lattice gauge theory for Haldane conjecture and
  central-branch Wilson fermion},''
  \href{http://dx.doi.org/10.1093/ptep/ptaa003}{{\em PTEP} {\bfseries 2020}
  no.~3, (2020) 033B03}, \href{http://arxiv.org/abs/1910.09604}{{\ttfamily
  arXiv:1910.09604 [hep-lat]}}.

\bibitem{Misumi:2020eyx}
T.~Misumi and J.~Yumoto, ``{Varieties and properties of central-branch Wilson
  fermions},'' \href{http://dx.doi.org/10.1103/PhysRevD.102.034516}{{\em Phys.
  Rev. D} {\bfseries 102} no.~3, (2020) 034516},
  \href{http://arxiv.org/abs/2005.08857}{{\ttfamily arXiv:2005.08857
  [hep-lat]}}.

\bibitem{Karsten:1981gd}
L.~H. Karsten, ``{Lattice Fermions in Euclidean Space-time},''
  \href{http://dx.doi.org/10.1016/0370-2693(81)90133-7}{{\em Phys. Lett. B}
  {\bfseries 104} (1981) 315--319}.

\bibitem{Wilczek:1987kw}
F.~Wilczek, ``{ON LATTICE FERMIONS},''
  \href{http://dx.doi.org/10.1103/PhysRevLett.59.2397}{{\em Phys. Rev. Lett.}
  {\bfseries 59} (1987) 2397}.

\bibitem{Creutz:2007af}
M.~Creutz, ``{Four-dimensional graphene and chiral fermions},''
  \href{http://dx.doi.org/10.1088/1126-6708/2008/04/017}{{\em JHEP} {\bfseries
  04} (2008) 017}, \href{http://arxiv.org/abs/0712.1201}{{\ttfamily
  arXiv:0712.1201 [hep-lat]}}.

\bibitem{Borici:2007kz}
A.~Borici, ``{Creutz fermions on an orthogonal lattice},''
  \href{http://dx.doi.org/10.1103/PhysRevD.78.074504}{{\em Phys. Rev. D}
  {\bfseries 78} (2008) 074504},
  \href{http://arxiv.org/abs/0712.4401}{{\ttfamily arXiv:0712.4401 [hep-lat]}}.

\bibitem{Bedaque:2008xs}
P.~F. Bedaque, M.~I. Buchoff, B.~C. Tiburzi, and A.~Walker-Loud, ``{Broken
  Symmetries from Minimally Doubled Fermions},''
  \href{http://dx.doi.org/10.1016/j.physletb.2008.03.034}{{\em Phys. Lett. B}
  {\bfseries 662} (2008) 449--455},
  \href{http://arxiv.org/abs/0801.3361}{{\ttfamily arXiv:0801.3361 [hep-lat]}}.

\bibitem{Bedaque:2008jm}
P.~F. Bedaque, M.~I. Buchoff, B.~C. Tiburzi, and A.~Walker-Loud, ``{Search for
  Fermion Actions on Hyperdiamond Lattices},''
  \href{http://dx.doi.org/10.1103/PhysRevD.78.017502}{{\em Phys. Rev. D}
  {\bfseries 78} (2008) 017502},
  \href{http://arxiv.org/abs/0804.1145}{{\ttfamily arXiv:0804.1145 [hep-lat]}}.

\bibitem{Capitani:2009yn}
S.~Capitani, J.~Weber, and H.~Wittig, ``{Minimally doubled fermions at one
  loop},'' \href{http://dx.doi.org/10.1016/j.physletb.2009.09.050}{{\em Phys.
  Lett. B} {\bfseries 681} (2009) 105--112},
  \href{http://arxiv.org/abs/0907.2825}{{\ttfamily arXiv:0907.2825 [hep-lat]}}.

\bibitem{Kimura:2009qe}
T.~Kimura and T.~Misumi, ``{Characters of Lattice Fermions Based on the
  Hyperdiamond Lattice},'' \href{http://dx.doi.org/10.1143/PTP.124.415}{{\em
  Prog. Theor. Phys.} {\bfseries 124} (2010) 415--432},
  \href{http://arxiv.org/abs/0907.1371}{{\ttfamily arXiv:0907.1371 [hep-lat]}}.

\bibitem{Kimura:2009di}
T.~Kimura and T.~Misumi, ``{Lattice Fermions Based on Higher-Dimensional
  Hyperdiamond Lattices},'' \href{http://dx.doi.org/10.1143/PTP.123.63}{{\em
  Prog. Theor. Phys.} {\bfseries 123} (2010) 63--78},
  \href{http://arxiv.org/abs/0907.3774}{{\ttfamily arXiv:0907.3774 [hep-lat]}}.

\bibitem{Creutz:2010cz}
M.~Creutz and T.~Misumi, ``{Classification of Minimally Doubled Fermions},''
  \href{http://dx.doi.org/10.1103/PhysRevD.82.074502}{{\em Phys. Rev. D}
  {\bfseries 82} (2010) 074502},
  \href{http://arxiv.org/abs/1007.3328}{{\ttfamily arXiv:1007.3328 [hep-lat]}}.

\bibitem{Creutz:2010qm}
M.~Creutz, ``{Minimal doubling and point splitting},''
  \href{http://dx.doi.org/10.22323/1.105.0078}{{\em PoS} {\bfseries
  LATTICE2010} (2010) 078}, \href{http://arxiv.org/abs/1009.3154}{{\ttfamily
  arXiv:1009.3154 [hep-lat]}}.

\bibitem{Capitani:2010nn}
S.~Capitani, M.~Creutz, J.~Weber, and H.~Wittig, ``{Renormalization of
  minimally doubled fermions},''
  \href{http://dx.doi.org/10.1007/JHEP09(2010)027}{{\em JHEP} {\bfseries 09}
  (2010) 027}, \href{http://arxiv.org/abs/1006.2009}{{\ttfamily arXiv:1006.2009
  [hep-lat]}}.

\bibitem{Tiburzi:2010bm}
B.~C. Tiburzi, ``{Chiral Lattice Fermions, Minimal Doubling, and the Axial
  Anomaly},'' \href{http://dx.doi.org/10.1103/PhysRevD.82.034511}{{\em Phys.
  Rev. D} {\bfseries 82} (2010) 034511},
  \href{http://arxiv.org/abs/1006.0172}{{\ttfamily arXiv:1006.0172 [hep-lat]}}.

\bibitem{Kamata:2011jn}
S.~Kamata and H.~Tanaka, ``{Minimal Doubling Fermion and Hermiticity},''
  \href{http://dx.doi.org/10.1093/ptep/pts093}{{\em PTEP} {\bfseries 2013}
  (2013) 023B05}, \href{http://arxiv.org/abs/1111.4536}{{\ttfamily
  arXiv:1111.4536 [hep-lat]}}.

\bibitem{Misumi:2012uu}
T.~Misumi, ``{Phase structure for lattice fermions with flavored chemical
  potential terms},'' \href{http://dx.doi.org/10.1007/JHEP08(2012)068}{{\em
  JHEP} {\bfseries 08} (2012) 068},
  \href{http://arxiv.org/abs/1206.0969}{{\ttfamily arXiv:1206.0969 [hep-lat]}}.

\bibitem{Misumi:2012ky}
T.~Misumi, T.~Kimura, and A.~Ohnishi, ``{QCD phase diagram with 2-flavor
  lattice fermion formulations},''
  \href{http://dx.doi.org/10.1103/PhysRevD.86.094505}{{\em Phys. Rev. D}
  {\bfseries 86} (2012) 094505},
  \href{http://arxiv.org/abs/1206.1977}{{\ttfamily arXiv:1206.1977 [hep-lat]}}.

\bibitem{Capitani:2013zta}
S.~Capitani, ``{Reducing the number of counterterms with new minimally doubled
  actions},'' \href{http://dx.doi.org/10.1103/PhysRevD.89.014501}{{\em Phys.
  Rev. D} {\bfseries 89} no.~1, (2014) 014501},
  \href{http://arxiv.org/abs/1307.7497}{{\ttfamily arXiv:1307.7497 [hep-lat]}}.

\bibitem{Capitani:2013iha}
S.~Capitani, ``{New chiral lattice actions of the Borici-Creutz type},''
  \href{http://dx.doi.org/10.1103/PhysRevD.89.074508}{{\em Phys. Rev. D}
  {\bfseries 89} no.~7, (2014) 074508},
  \href{http://arxiv.org/abs/1311.5664}{{\ttfamily arXiv:1311.5664 [hep-lat]}}.

\bibitem{Misumi:2013maa}
T.~Misumi, ``{Fermion Actions extracted from Lattice Super Yang-Mills
  Theories},'' \href{http://dx.doi.org/10.1007/JHEP12(2013)063}{{\em JHEP}
  {\bfseries 12} (2013) 063}, \href{http://arxiv.org/abs/1311.4365}{{\ttfamily
  arXiv:1311.4365 [hep-lat]}}.

\bibitem{Weber:2013tfa}
J.~H. Weber, S.~Capitani, and H.~Wittig, ``{Numerical studies of Minimally
  Doubled Fermions},'' \href{http://dx.doi.org/10.22323/1.187.0122}{{\em PoS}
  {\bfseries LATTICE2013} (2014) 122},
  \href{http://arxiv.org/abs/1312.0488}{{\ttfamily arXiv:1312.0488 [hep-lat]}}.

\bibitem{Weber:2017eds}
J.~H. Weber, {\em {Properties of minimally doubled fermions}}.
\newblock PhD thesis, Mainz U., 2015.
\newblock \href{http://arxiv.org/abs/1706.07104}{{\ttfamily arXiv:1706.07104
  [hep-lat]}}.

\bibitem{Durr:2020yqa}
S.~D\"urr and J.~H. Weber, ``{Dispersion relation and spectral range of
  Karsten-Wilczek and Borici-Creutz fermions},''
  \href{http://dx.doi.org/10.1103/PhysRevD.102.014516}{{\em Phys. Rev. D}
  {\bfseries 102} (Jul, 2020) 014516},
  \href{http://arxiv.org/abs/2003.10803}{{\ttfamily arXiv:2003.10803
  [hep-lat]}}. \url{https://link.aps.org/doi/10.1103/PhysRevD.102.014516}.

\bibitem{Wan:2010fyf}
X.~Wan, A.~Turner, A.~Vishwanath, and S.~Y. Savrasov, ``{Topological semimetal
  and Fermi-arc surface states in the electronic structure of pyrochlore
  iridates},'' \href{http://dx.doi.org/10.1103/PhysRevB.83.205101}{{\em Phys.
  Rev. B} {\bfseries 83} no.~20, (2011) 205101},
  \href{http://arxiv.org/abs/1007.0016}{{\ttfamily arXiv:1007.0016
  [cond-mat.str-el]}}.

\bibitem{Gioia:2025bhl}
L.~Gioia and R.~Thorngren, ``{Exact Chiral Symmetries of 3+1D Hamiltonian
  Lattice Fermions},'' \href{http://arxiv.org/abs/2503.07708}{{\ttfamily
  arXiv:2503.07708 [cond-mat.str-el]}}.

\bibitem{Creutz:2001wp}
M.~Creutz, I.~Horvath, and H.~Neuberger, ``{A New fermion Hamiltonian for
  lattice gauge theory},''
  \href{http://dx.doi.org/10.1016/S0920-5632(01)01836-9}{{\em Nucl. Phys. B
  Proc. Suppl.} {\bfseries 106} (2002) 760--762},
  \href{http://arxiv.org/abs/hep-lat/0110009}{{\ttfamily
  arXiv:hep-lat/0110009}}.

\bibitem{Fujikawa:2000my}
K.~Fujikawa, ``{Algebraic generalization of the Ginsparg-Wilson relation},''
  \href{http://dx.doi.org/10.1016/S0550-3213(00)00395-3}{{\em Nucl. Phys. B}
  {\bfseries 589} (2000) 487--503},
  \href{http://arxiv.org/abs/hep-lat/0004012}{{\ttfamily
  arXiv:hep-lat/0004012}}.

\bibitem{Clancy:2023kla}
M.~Clancy, ``{Toward a Ginsparg-Wilson lattice Hamiltonian},''
  \href{http://dx.doi.org/10.1103/PhysRevD.110.L011502}{{\em Phys. Rev. D}
  {\bfseries 110} no.~1, (2024) L011502},
  \href{http://arxiv.org/abs/2312.08647}{{\ttfamily arXiv:2312.08647
  [hep-lat]}}.

\bibitem{Clancy:2023ino}
M.~Clancy, D.~B. Kaplan, and H.~Singh, ``{Generalized Ginsparg-Wilson
  relations},'' \href{http://dx.doi.org/10.1103/PhysRevD.109.014502}{{\em Phys.
  Rev. D} {\bfseries 109} no.~1, (2024) 014502},
  \href{http://arxiv.org/abs/2309.08542}{{\ttfamily arXiv:2309.08542
  [hep-lat]}}.

\bibitem{Fidkowski:2023sif}
L.~Fidkowski and C.~Xu, ``{A No-Go Result for Implementing Chiral Symmetries by
  Locality-Preserving Unitaries in a Three-Dimensional Hamiltonian Lattice
  Model of Fermions},''
  \href{http://dx.doi.org/10.1103/PhysRevLett.131.196601}{{\em Phys. Rev.
  Lett.} {\bfseries 131} no.~19, (2023) 196601},
  \href{http://arxiv.org/abs/2306.10105}{{\ttfamily arXiv:2306.10105
  [cond-mat.str-el]}}.

\bibitem{Singh:2025sye}
H.~Singh, ``{Ginsparg-Wilson Hamiltonians with Improved Chiral Symmetry},''
  \href{http://arxiv.org/abs/2505.20419}{{\ttfamily arXiv:2505.20419
  [hep-lat]}}.

\bibitem{Meyniel:2025euu}
G.~Meyniel and F.~Zhou, ``{Equivalent class of Emergent Single Weyl Fermion in
  3d Topological States: gapless superconductors and superfluids Vs chiral
  fermions},'' \href{http://arxiv.org/abs/2510.25959}{{\ttfamily
  arXiv:2510.25959 [hep-th]}}.

\bibitem{Chatterjee:2024gje}
A.~Chatterjee, S.~D. Pace, and S.-H. Shao, ``{Quantized Axial Charge of
  Staggered Fermions and the Chiral Anomaly},''
  \href{http://dx.doi.org/10.1103/PhysRevLett.134.021601}{{\em Phys. Rev.
  Lett.} {\bfseries 134} no.~2, (2025) 021601},
  \href{http://arxiv.org/abs/2409.12220}{{\ttfamily arXiv:2409.12220
  [hep-th]}}.

\bibitem{Onogi:2025xir}
T.~Onogi and T.~Yamaoka, ``{Non-singlet conserved charges and anomalies in 3+1
  D staggered fermions},'' \href{http://arxiv.org/abs/2509.04906}{{\ttfamily
  arXiv:2509.04906 [hep-lat]}}.

\bibitem{Aoki:2025vtp}
S.~Aoki, Y.~Kikukawa, and T.~Takemoto, ``{Chiral Anomaly of Kogut-Susskind
  Fermion in (3+1)-dimensional Hamiltonian formalism},''
  \href{http://arxiv.org/abs/2511.06198}{{\ttfamily arXiv:2511.06198
  [hep-lat]}}.

\bibitem{Eichten:1985ft}
E.~Eichten and J.~Preskill, ``{Chiral Gauge Theories on the Lattice},''
\href{http://dx.doi.org/10.1016/0550-3213(86)90207-5}{{\em Nucl. Phys.}
  {\bfseries B268} (1986) 179--208}.

\bibitem{Sint:2007ug}
S.~Sint, \href{http://dx.doi.org/10.1142/9789812790927\_0004}{``{Lattice QCD
  with a chiral twist},''} in {\em {Workshop on Perspectives in Lattice QCD}}.
\newblock 2, 2007.
\newblock \href{http://arxiv.org/abs/hep-lat/0702008}{{\ttfamily
  arXiv:hep-lat/0702008}}.

\bibitem{Shindler:2007vp}
A.~Shindler, ``{Twisted mass lattice QCD},''
  \href{http://dx.doi.org/10.1016/j.physrep.2008.03.001}{{\em Phys. Rept.}
  {\bfseries 461} (2008) 37--110},
  \href{http://arxiv.org/abs/0707.4093}{{\ttfamily arXiv:0707.4093 [hep-lat]}}.

\bibitem{Shukre:2025bnm}
K.~Shukre and S.~Basak, ``{Symanzik effective action for Karsten-Wilczek
  minimally doubled fermions},''
  \href{http://dx.doi.org/10.1103/l6k8-pgqg}{{\em Phys. Rev. D} {\bfseries 112}
  no.~11, (2025) 114501}, \href{http://arxiv.org/abs/2508.09690}{{\ttfamily
  arXiv:2508.09690 [hep-lat]}}.

\bibitem{Weber:2025kcl}
J.~H. Weber, ``{Spin-taste representation of minimally doubled fermions from
  first principles: Karsten-Wilczek fermions},''
  \href{http://dx.doi.org/10.1103/PhysRevD.112.014509}{{\em Phys. Rev. D}
  {\bfseries 112} no.~1, (2025) 014509},
  \href{http://arxiv.org/abs/2502.16500}{{\ttfamily arXiv:2502.16500
  [hep-lat]}}.

\bibitem{Shukre:2024tkw}
K.~S. Shukre, D.~Chakrabarti, and S.~Basak, ``{Chiral Lagrangian for
  Karsten-Wilczek Minimally Doubled Fermion},''
  \href{http://dx.doi.org/10.22323/1.466.0357}{{\em PoS} {\bfseries
  LATTICE2024} (2025) 357}, \href{http://arxiv.org/abs/2411.14848}{{\ttfamily
  arXiv:2411.14848 [hep-lat]}}.

\bibitem{Weber:2023kth}
J.~H. Weber, ``{Spin-taste structure of minimally doubled fermions},''
  \href{http://dx.doi.org/10.22323/1.453.0353}{{\em PoS} {\bfseries
  LATTICE2023} (2024) 353}, \href{http://arxiv.org/abs/2312.08526}{{\ttfamily
  arXiv:2312.08526 [hep-lat]}}.

\bibitem{Borsanyi:2025big}
S.~Bors{\'a}nyi, S.~Capitani, Z.~Fodor, D.~Godzieba, P.~Parotto, R.~A. Vig, and
  C.~H. Wong, ``{Taste breaking in the minimally doubled Karsten-Wilczek action
  and its tree-level improvement},''
  \href{http://dx.doi.org/10.1103/PhysRevD.111.074510}{{\em Phys. Rev. D}
  {\bfseries 111} no.~7, (2025) 074510},
  \href{http://arxiv.org/abs/2502.07354}{{\ttfamily arXiv:2502.07354
  [hep-lat]}}.

\bibitem{Vig:2024umj}
R.~A. Vig, S.~Borsanyi, Z.~Fodor, D.~Godzieba, P.~Parotto, and C.~H. Wong,
  ``{First dynamical simulations with minimally doubled fermions},''
  \href{http://dx.doi.org/10.22323/1.453.0289}{{\em PoS} {\bfseries
  LATTICE2023} (2024) 289}, \href{http://arxiv.org/abs/2401.07651}{{\ttfamily
  arXiv:2401.07651 [hep-lat]}}.

\bibitem{Durr:2024ttb}
S.~Durr and S.~Capitani, ``{Taste-splittings of staggered, Karsten-Wilczek and
  Borici-Creutz fermions under gradient flow in 2D},''
  \href{http://dx.doi.org/10.22323/1.466.0460}{{\em PoS} {\bfseries
  LATTICE2024} (2025) 460}, \href{http://arxiv.org/abs/2411.18237}{{\ttfamily
  arXiv:2411.18237 [hep-lat]}}.

\end{thebibliography}\endgroup

\end{document}